\documentclass[twocolumn]{aastex62}

\usepackage{multirow}
\usepackage{graphicx}

\received{1 December 2019}
\revised{21 April 2020}
\accepted{17 May 2020}

\submitjournal{AJ}

\begin{document}

\title{ARES\footnote{ARES: Ariel Retrieval of Exoplanets School} II: Characterising the Hot Jupiters WASP-127\,b, WASP-79\,b and WASP-62\,b with HST}

\correspondingauthor{Nour Skaf}
\email{nour.skaf@obspm.fr}

 \author{Nour Skaf} 
 	\affil{LESIA, Observatoire de Paris, Université PSL, CNRS, Sorbonne Universit\'e, Universit\'e de Paris, 5 place Jules Janssen, 92195 Meudon, France}
	 \affil{Department of Physics and Astronomy, University College London, London, United Kingdom}

\author{Michelle Fabienne Bieger}
\affil{College of Engineering, Mathematics and Physical Sciences,  University of Exeter, North Park Road, Exeter, United Kingdom}

\author{Billy Edwards}
\affil{Department of Physics and Astronomy, University College London, London, United Kingdom}

\author{Quentin Changeat}
\affil{Department of Physics and Astronomy, University College London, London, United Kingdom}

\author{Mario Morvan}
	\affil{Department of Physics and Astronomy, University College London, London, United Kingdom}

\author{Flavien Kiefer}
\affil{Sorbonne Universit\'es, UPMC Universit\'e Paris 6 et CNRS, 
UMR 7095, Institut d'Astrophysique de Paris, 98 bis bd Arago,
75014 Paris, France}

\author{Doriann Blain}
 	\affil{LESIA, Observatoire de Paris, Université PSL, CNRS, Sorbonne Universit\'e, Universit\'e de Paris, 5 place Jules Janssen, 92195 Meudon, France}

 \author{Tiziano Zingales} 
 	\affil{Laboratoire d'astrophysique de Bordeaux, Univ. Bordeaux, CNRS, B18N, all\'{e}e Geoffroy Saint-Hilaire, 33615 Pessac, France}

\author{Mathilde Poveda}
	\affil{Laboratoire Interuniversitaire des Syst\`{e}mes Atmosph\'{e}riques (LISA), UMR CNRS 7583, Universit\'{e} Paris-Est-Cr\'eteil, Universit\'e de Paris, Institut Pierre Simon Laplace, Cr\'{e}teil, France}
	\affil{Maison de la Simulation, CEA, CNRS, Univ. Paris-Sud, UVSQ, Universit\'{e} Paris-Saclay, F-91191 Gif-sur-Yvette, France} 

\author{Ahmed Al-Refaie} 
\affil{Department of Physics and Astronomy, University College London, London, United Kingdom}

\author{Robin Baeyens}
\affil{Instituut voor Sterrenkunde, KU Leuven, Celestijnenlaan 200D bus 2401, 3001 Leuven, Belgium}

\author{Am\'elie Gressier}
\affil{LATMOS, CNRS, Sorbonne Universit\'e UVSQ, 11 boulevard d’Alembert, F-78280 Guyancourt, France}
\affil{Sorbonne Universit\'es, UPMC Universit\'e Paris 6 et CNRS, 
UMR 7095, Institut d'Astrophysique de Paris, 98 bis bd Arago,
75014 Paris, France}
 	\affil{LESIA, Observatoire de Paris, Université PSL, CNRS, Sorbonne Universit\'e, Universit\'e de Paris, 5 place Jules Janssen, 92195 Meudon, France}

\author{Gloria Guilluy}
	\affil{Dipartimento di Fisica, Università degli Studi di Torino, via Pietro Giuria 1, I-10125 Torino, Italy}
	\affil{INAF - Osservatorio Astrofisico di Torino, Via Osservatorio 20, I-10025 Pino Torinese, Italy}

\author{Adam Yassin Jaziri}  
\affil{Laboratoire d'astrophysique de Bordeaux, Univ. Bordeaux, CNRS, B18N, all\'{e}e Geoffroy Saint-Hilaire, 33615 Pessac, France}

\author{Darius Modirrousta-Galian} 
	\affil{INAF - Osservatorio Astronomico di Palermo, Piazza del Parlamento 1, I-90134 Palermo, Italy}
	\affil{University of Palermo, Department of Physics and Chemistry, Via Archirafi 36, Palermo, Italy}

\author{Lorenzo V. Mugnai} 
	\affil{La Sapienza Universit\'a di Roma, Department of Physics, Piazzale Aldo Moro 2, 00185 Roma, Italy}

\author{William Pluriel}  
	\affil{Laboratoire d'astrophysique de Bordeaux, Univ. Bordeaux, CNRS, B18N, all\'{e}e Geoffroy Saint-Hilaire, 33615 Pessac, France}

\author{Niall Whiteford}
 \affil{Institute for Astronomy, University of Edinburgh, Blackford Hill, Edinburgh EH9 3HJ, UK}
    \affil{Centre for Exoplanet Science, University of Edinburgh, Edinburgh EH9 3FD, UK}

\author{Sam Wright}
\affil{Department of Physics and Astronomy, University College London, London, United Kingdom}

\author{Kai Hou Yip}
\affil{Department of Physics and Astronomy, University College London, London, United Kingdom}

\author{Benjamin Charnay}
 	\affil{LESIA, Observatoire de Paris, Université PSL, CNRS, Sorbonne Universit\'e, Universit\'e de Paris, 5 place Jules Janssen, 92195 Meudon, France}

\author{J\'{e}r\'{e}my Leconte}  
\affil{Laboratoire d'astrophysique de Bordeaux, Univ. Bordeaux, CNRS, B18N, all\'{e}e Geoffroy Saint-Hilaire, 33615 Pessac, France}

\author{Pierre Drossart}  
 	\affil{LESIA, Observatoire de Paris, Université PSL, CNRS, Sorbonne Universit\'e, Universit\'e de Paris, 5 place Jules Janssen, 92195 Meudon, France}

\author{Angelos Tsiaras}
\affil{Department of Physics and Astronomy, University College London, London, United Kingdom}

\author{Olivia Venot}  
\affil{Laboratoire Interuniversitaire des Syst\`{e}mes Atmosph\'{e}riques (LISA), UMR CNRS 7583, Universit\'{e} Paris-Est-Cr\'eteil, Universit\'e de Paris, Institut Pierre Simon Laplace, Cr\'{e}teil, France}

\author{Ingo Waldmann}
\affil{Department of Physics and Astronomy, University College London, London, United Kingdom}

\author{Jean-Philippe Beaulieu}
\affil{School of Physical Sciences, University of Tasmania,
Private Bag 37 Hobart, Tasmania 7001 Australia}
\affil{Sorbonne Universit\'es, UPMC Universit\'e Paris 6 et CNRS, 
UMR 7095, Institut d'Astrophysique de Paris, 98 bis bd Arago,
75014 Paris, France}

\begin{abstract}

This paper presents the atmospheric characterisation of three large, gaseous planets: WASP-127\,b, WASP-79\,b and WASP-62\,b. We analysed spectroscopic data obtained with the G141 grism (1.088 - 1.68 $\mu$m) of the Wide Field Camera 3 (WFC3) onboard the Hubble Space Telescope (HST) using the \verb+Iraclis+ pipeline and the TauREx3 retrieval code, both of which are publicly available. For WASP-127\,b, which is the least dense planet discovered so far and is located in the short-period Neptune desert, our retrieval results found strong water absorption corresponding to an abundance of log(H$_2$O) = -2.71$^{+0.78}_{-1.05}$, and absorption compatible with an iron hydride abundance of log(FeH)=$-5.25^{+0.88}_{-1.10}$, with an extended cloudy atmosphere.
We also detected water vapour in the atmospheres of WASP-79\,b and WASP-62\,b, with best-fit models indicating the presence of iron hydride, too.
We used the Atmospheric Detectability Index (ADI) as well as Bayesian log evidence to quantify the strength of the detection and compared our results to the hot Jupiter population study by \cite{angelos30}.
While all the planets studied here are suitable targets for characterisation with upcoming facilities such as the James Webb Space Telescope (JWST) and Ariel, WASP-127\,b is of particular interest due to its low density, and a thorough atmospheric study would develop our understanding of planet formation and migration. 

\end{abstract}


\section{Introduction} \label{sec:intro}

The currently-known exoplanet population displays a wide range of masses, radii, and orbits. Although many planets have been detected and it is thought that they are common in our Galaxy \citep{howard,batalha,cassan,dressing,wright_jup}, our current knowledge of their atmospheric characteristics is still very limited. Examining the atmospheres of exoplanets further unveils their planetary properties, with their study made possible by various methods, including transit spectroscopy \citep[e.g.][]{Tinetti_water, mark}. Facilities such as the  \textit{Hubble Space Telescope} and the \textit{Spitzer Space Telescope}, as well as some ground-based observatories, have provided constraints on these properties for a limited number of targets and, in some cases, have identified the key molecules present in their atmospheres while also detecting the presence of clouds and probing their thermal structure \citep[e.g.][]{brogi,majeau,stevenson,sing,fu,angelos30,kelt9_iron,pinhas,k2_18b,iron_wasp76b,ares1}.

This paper presents the analysis of data from Hubble's public archive for the exoplanets WASP-127\,b \citep{lam_wasp127}, WASP-79\,b \citep{smalley12} and WASP-62\,b \citep{Hellier2012}. They are all inflated, with low eccentricities and short orbital periods around bright stars. Table \ref{tab:general_parameters} presents the stellar and planetary parameters for each of these targets. 

WASP-127\,b is an ideal target for spectroscopic studies, given its unusually low density (with a super-Jupiter radius and a sub-Saturn mass). It is located in the short-period Neptune desert, where it is expected that planets might not survive photo-evaporation \citep{Owen, Mazeh}. However, photo-evaporation is strongly case-dependent and this planet receives a relatively low extreme ultraviolet flux \citep{Chen}. Potential explanations for its inflation include tidal heating, enhanced atmospheric opacity, Ohmic heating, and re-inflation by the host star during inward migration \citep{jeremy2010,2010ohm,2011ohm,rauscher_menou,2014wu}. 
Both WASP-62\,b and WASP-79\,b are believed to have an evaporating atmosphere, with mass loss rates estimated at $\approx$ 11 g$\cdot$s$^{-1}$ \citep{Bourrier2015}. WASP-79\,b, which has a polar orbit, was originally detected through an aberration in the radial velocity due to the Rossiter McLaughlin effect \citep{Addison13}.

All spectral data presented herein were acquired with the G141 grism (1.088 - 1.68 $\mu$m) of the HST/WFC3 camera and details regarding each observation can be found in Table \ref{tab:HSTdata}. In Section \ref{sec:dataanalysis}, we detail how the data were reduced with the \verb+Iraclis+ pipeline \citep{Iraclis}, following the approach described by \citet{angelos30}, and summarised here. In Section \ref{sec:atmochar}, we describe the TauREx retrieval code used to analyse the reduced spectra \citep{Waldmann2015b, Waldmann2015a, al-refaie_taurex3}, along with the initial parameters and priors used. Our results can be found in Section \ref{sec:results}, followed by a discussion on our findings and the implications they hold for future missions, including simulations of data from Ariel and JWST.

\begin{table*}[ht!]
\caption{\label{tab:general_parameters} Target Parameters}
\centering
\begin{tabular}{@{}lccc@{}} \hline\hline
Parameter & WASP-127\,b & WASP-79\,b &  WASP-62\,b \\ \hline
\multicolumn{4}{c}{Stellar parameters} \\ \hline
Spectral type & G5 & F5 & F7 \\
$T_{\text{eff}}$ [K] & 5750$\pm$85 & 6600$\pm$100 &  6230$\pm$80 \\
$\log g$ (cgs) & 3.9  & 4.06$\pm$0.15 &  4.45$\pm$0.10\\
$[$Fe/H$]$ & -0.18$\pm$0.06 & 0.03 & 0.04  \\ \hline
\multicolumn{4}{c}{Planetary parameters} \\ \hline
$P$ [days] &  4.17807015$\pm$2.10$^{-6}$ & 3.662387$\pm$4.10$^{-6}$ &   4.411953$\pm$4.10$^{-6}$ \\
$T_{mid}$ [BJD$_{TDB}$ - 2450000] & 8138.670144 & 7815.89868 & 5855.39195\\
$i$ [$^\circ$] & 88.2$^{+1.1}_{-0.9}$ & 86.1$\pm$0.2 &  88.5 $^{+0.4}_{-0.7}$\\
$M_{\text{P}} [M_J]$ & 0.18$\pm$0.02 & 0.85$\pm$0.8 & 0.58$\pm$0.03 \\
$R_{\text{P}} [R_J]$ & 1.37$\pm$0.04  & 1.53$\pm$0.04 & 1.34$^{+0.05}_{-0.03} $ \\
$T_{\text{eq},A=0}$ [K] & 1400$\pm$24 & 1716.2$^{+25.8}_{-24.4}$ & 1475.3$^{+25.1}_{-10}$ \\ 
$R_{P}$/$R_\star$ & 0.09992$^{+0.0028}_{-0.0029}$   & 0.09609$^{+0.0023}_{-0.0027}$ & 0.1091$^{+0.0038}_{-0.0023}$ \\
$a$/$R_\star$ & 7.846 & 6.069 & 9.5253 \\ \hline
References & \cite{Palle} & \cite{brown_rm} & \cite{brown_rm} \\
\hline \hline
\end{tabular}
\end{table*}


\begin{table*}[ht!]\centering
\caption{\label{tab:HSTdata} HST/WFC3 data summary.}
\begin{tabular}{lccccc}
\hline
Planet & Median epoch & Mean exposure time  & Number of spectra &  PI name & Proposal ID \\ 
       & (MJD)  & (sec)  \\
\hline
WASP-127\,b & 58217.51310 & 95.782    & 74 &  Jessica Spake & 14619 \\
WASP-79\,b  & 57815.37216 & 138.381 & 64 &  David Sing  & 14767  \\
WASP-62\,b  & 57857.82823 &  138.381 & 61 & David Sing  & 14767  \\
\hline
\end{tabular}
\end{table*}

\section{Methodology}\label{sec:dataanalysis}
\subsection{HST Observations}
Data reduction and calibration were performed using \verb+Iraclis+, software developed in \citet{Iraclis} and available on GitHub\footnote{\url{https://github.com/ucl-exoplanets/Iraclis}}; it has been used to extract HST spectra in multiple studies, including \citet{55cancri,angelos30,k2_18b}. We used the Mikulski Archive for Space Telescopes\footnote{\url{https://archive.stsci.edu/hst/}} to assess the spectroscopic observational data of WASP-127\,b, WASP-79\,b and WASP-62\,b and information about the observations can be found in Table \ref{tab:HSTdata}. The WASP-127\,b proposal was led by Jessica Spake while David Sing was the PI for the observations of WASP-79\,b and WASP-62\,b. Although more data is available from additional instruments, we have restricted our study to HST/WFC3 data, in order to maintain consistency in comparing the analysis of the planets. 

\subsection{Data Analysis}
The planets in this paper were analysed to be comparable to the thirty planets studied in \citet{angelos30}. We followed the same methodology as summarised below; differences between our studies are stated explicitly. 

Our analysis began with raw spatially scanned spectroscopic images, with data reduction and correction steps performed in the following order: zero-read subtraction; reference pixel correction; non-linearity correction; dark current subtraction; gain conversion; sky background subtraction; calibration; flat-field correction; bad pixels and cosmic ray correction. 

Following the reduction process, the flux was extracted from the spatially scanned spectroscopic images to create the final transit light-curves per wavelength band. We considered one broadband (white) light-curve covering the whole wavelength range in which the G141 grism is sensitive (1.088 - 1.68 $\mu$m) and spectral light-curves with a resolving power of 70 at 1.4 $\mu$m. When extracting the spectral light-curves, \verb+Iraclis+ accounts for the geometric distortions induced by the tilt of the detector in the WFC3 infrared channel. The bands of the spectral light-curves are selected such that the SNR is approximately uniform across the planetary spectrum. We extracted our final light-curves from the differential, non-destructive reads. Prior to light curve fitting, we chose to discard the first HST orbit of each visit, as these exhibit much stronger hooks than subsequent orbits.

\begin{figure}[ht!]
    \centering
    \includegraphics[width=\columnwidth]{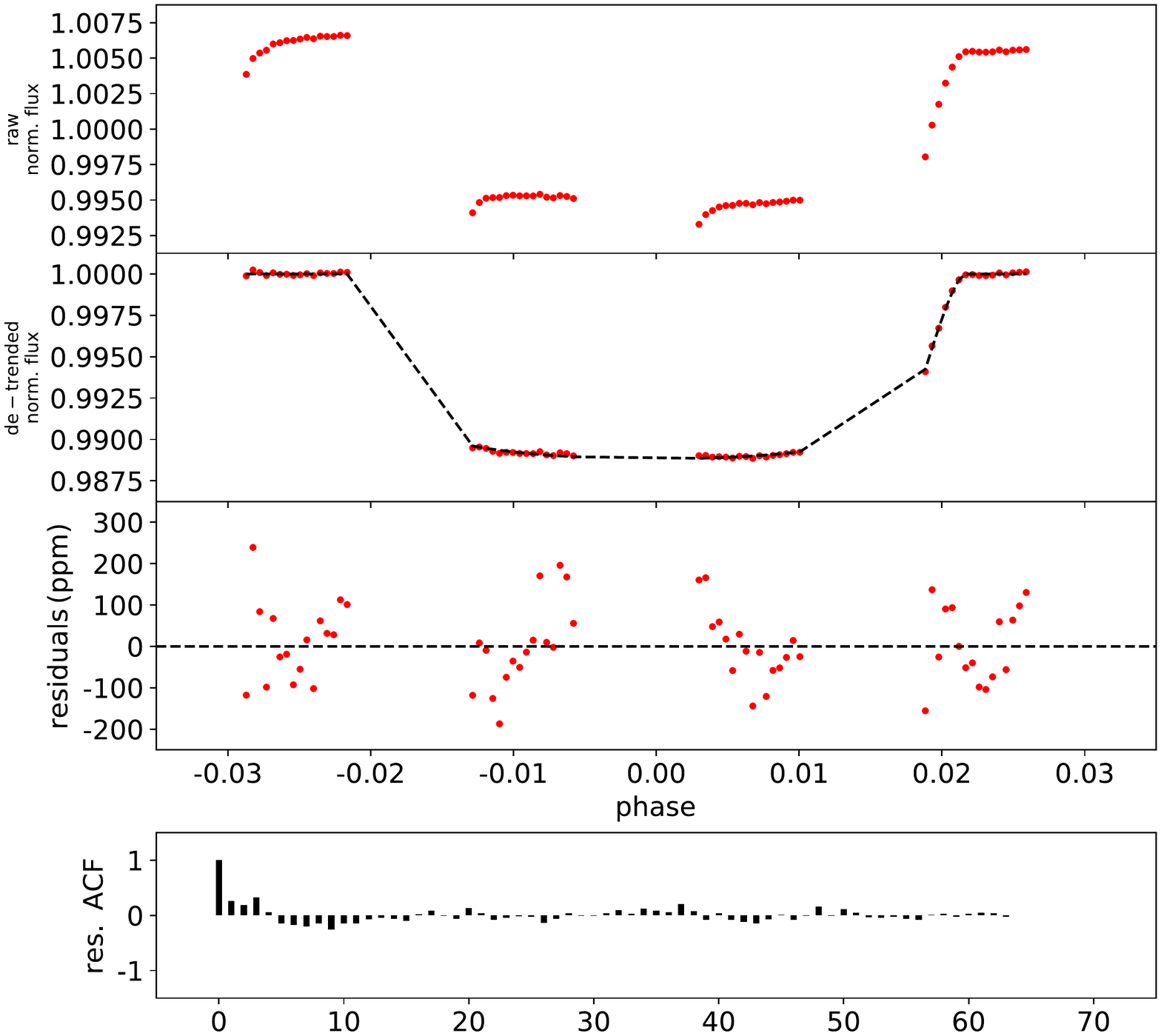}
    \caption{Results of the white light-curve of WASP-127\,b. Top: raw light-curve, after normalisation. Second: light-curve, divided by the best fit model for the systematics. Third: residuals. Bottom: auto-correlation function of the residuals.}
    \label{fig:white}
\end{figure}

\begin{figure}[ht!]
    \centering
    \includegraphics[width=\columnwidth]{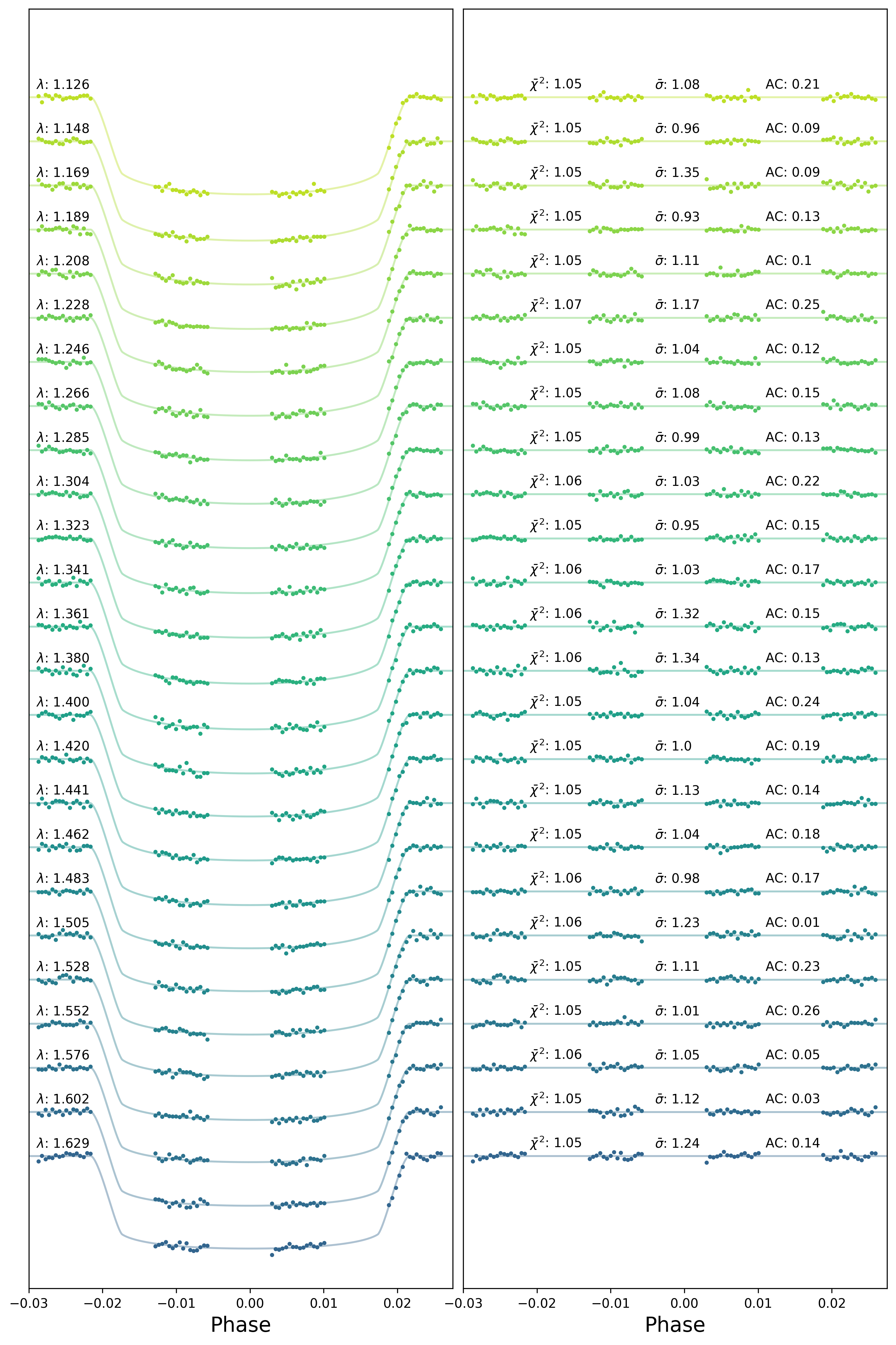}
    \caption{Spectral light curves fitted with Iraclis for the transmission spectra where, for clarity, an offset has been applied. Left: the detrended spectral light curves with best-fit model plotted. Right: residuals from the fitting with values for the Chi-squared ($\chi^2$), the standard deviation of the residuals with respect to the photon noise ($\bar{\sigma}$) and the auto-correlation (AC). The mean $\bar{\sigma}$ for each of the three planets is between 1.02 and 1.25 times the photon noise.}
    \label{fig:all}
\end{figure}

\begin{figure}
\centering
\includegraphics[width=\columnwidth]{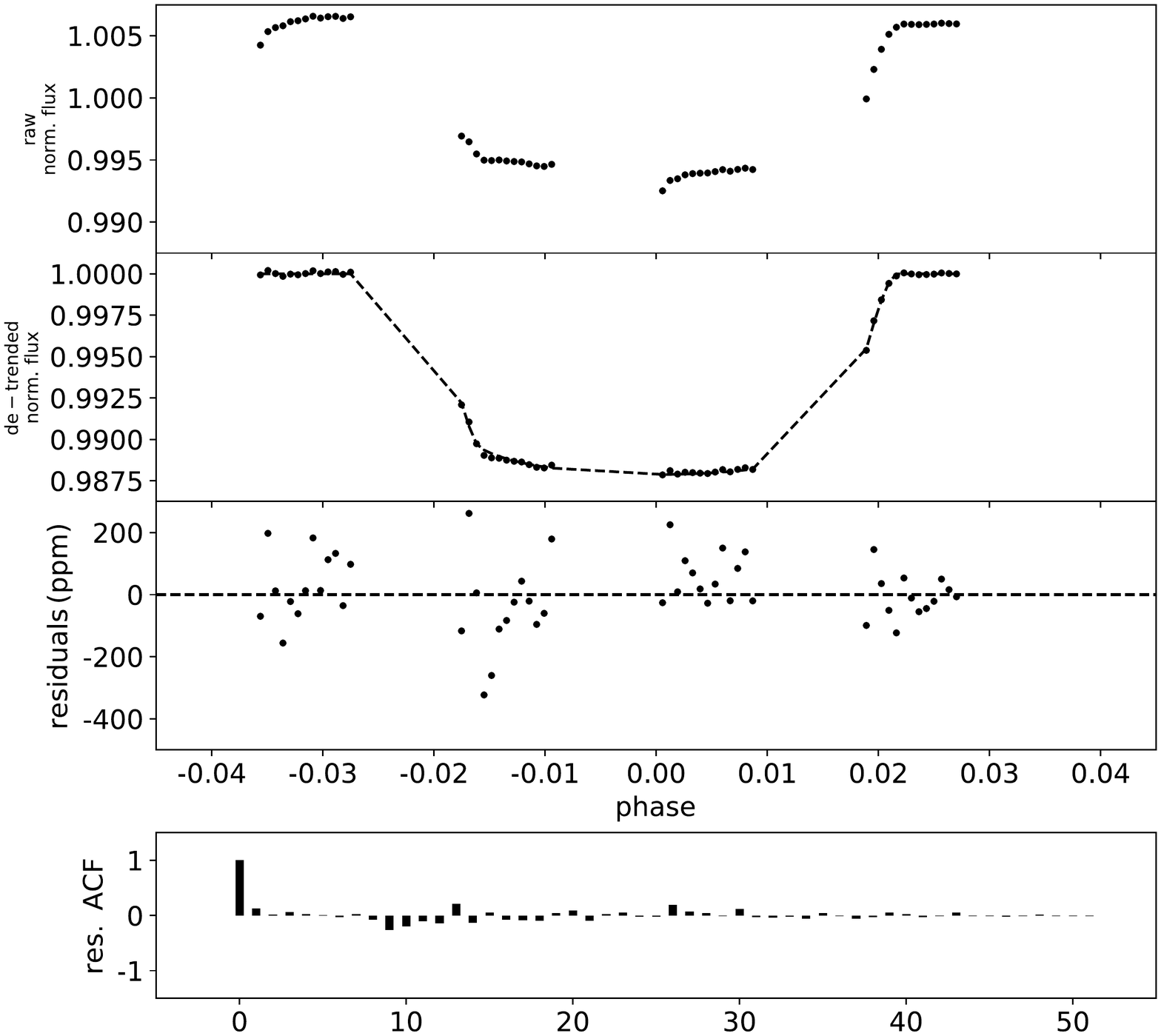}
\includegraphics[width=0.95\columnwidth]{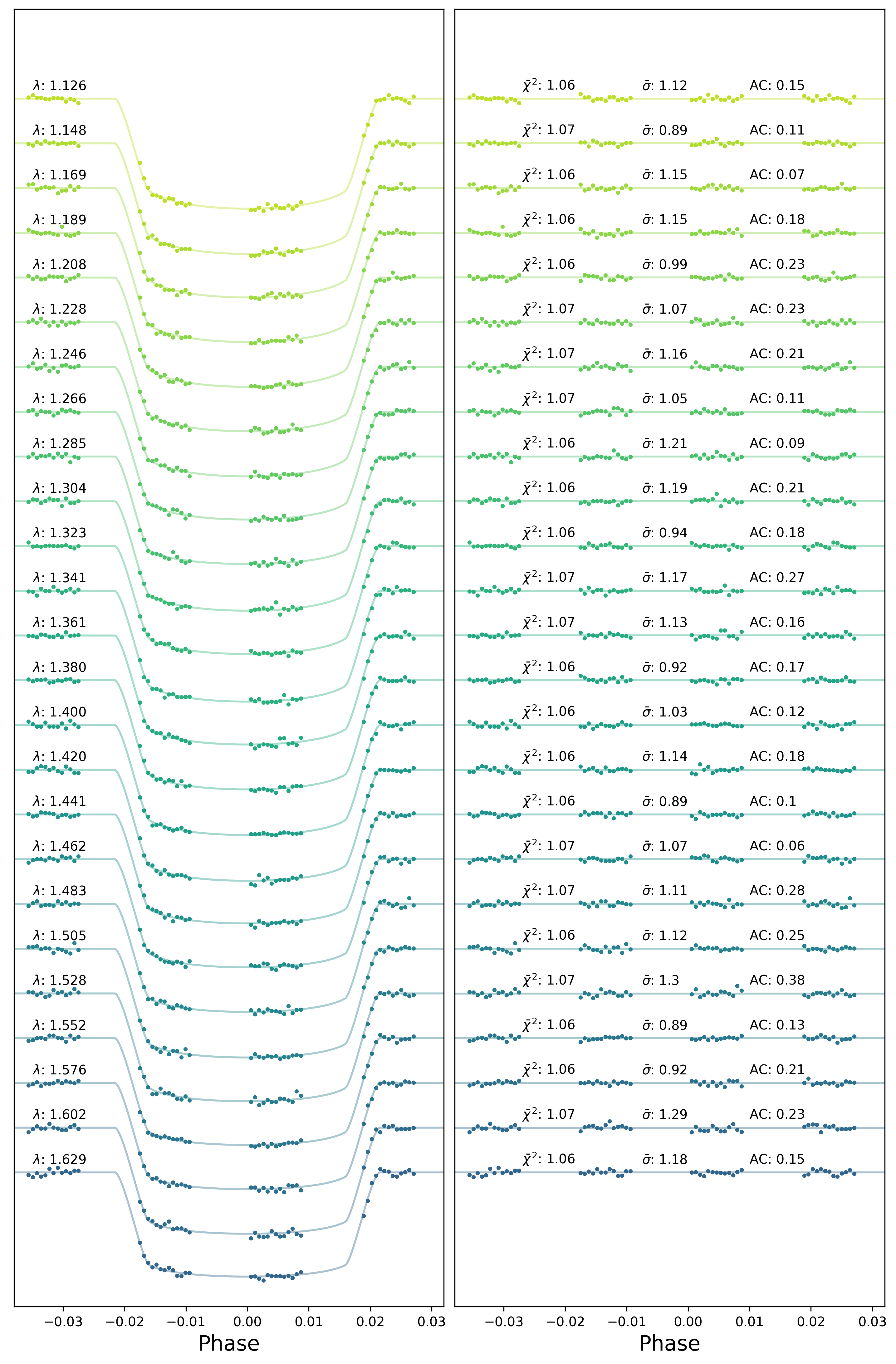}
\caption{Top figure: Results of the white light-curve of WASP-79\,b. Top: raw light-curve, after normalisation. Second: light-curve, divided by the best fit model for the systematics. Third: residuals. Bottom: auto-correlation function of the residuals. Bottom figure: Spectral light curves fitted with Iraclis for the transmission spectra where, for clarity, an offset has been applied. Left: the detrended spectral light curves with best-fit model plotted. Right: residuals from the fitting with values for the Chi-squared ($\chi^2$), the standard deviation of the residuals with respect to the photon noise ($\bar{\sigma}$) and the auto-correlation (AC).}
\label{fig:wasp79_lc}
\end{figure}

\begin{figure}
\centering
\includegraphics[width=0.8\columnwidth]{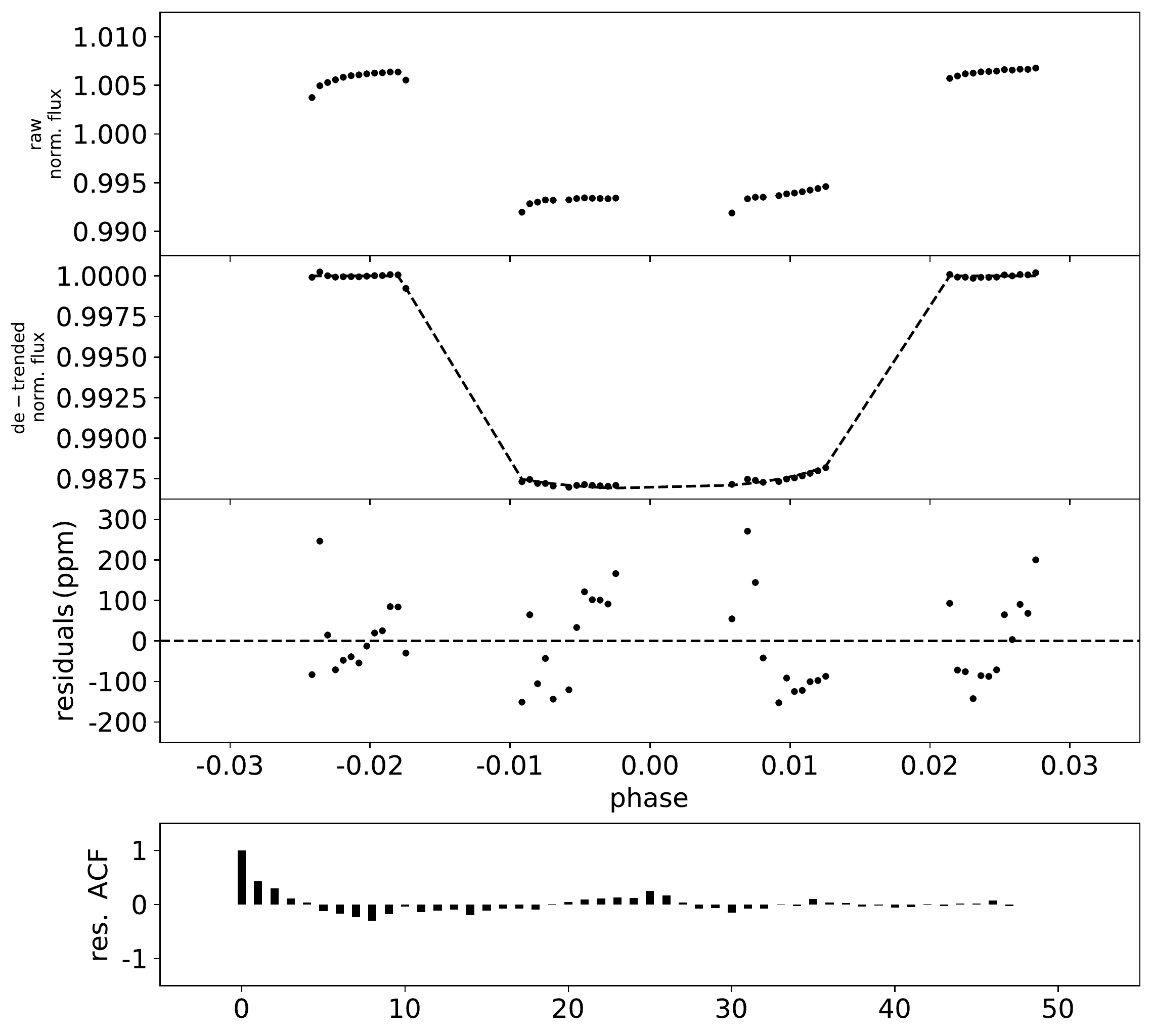}
\includegraphics[width=0.95\columnwidth]{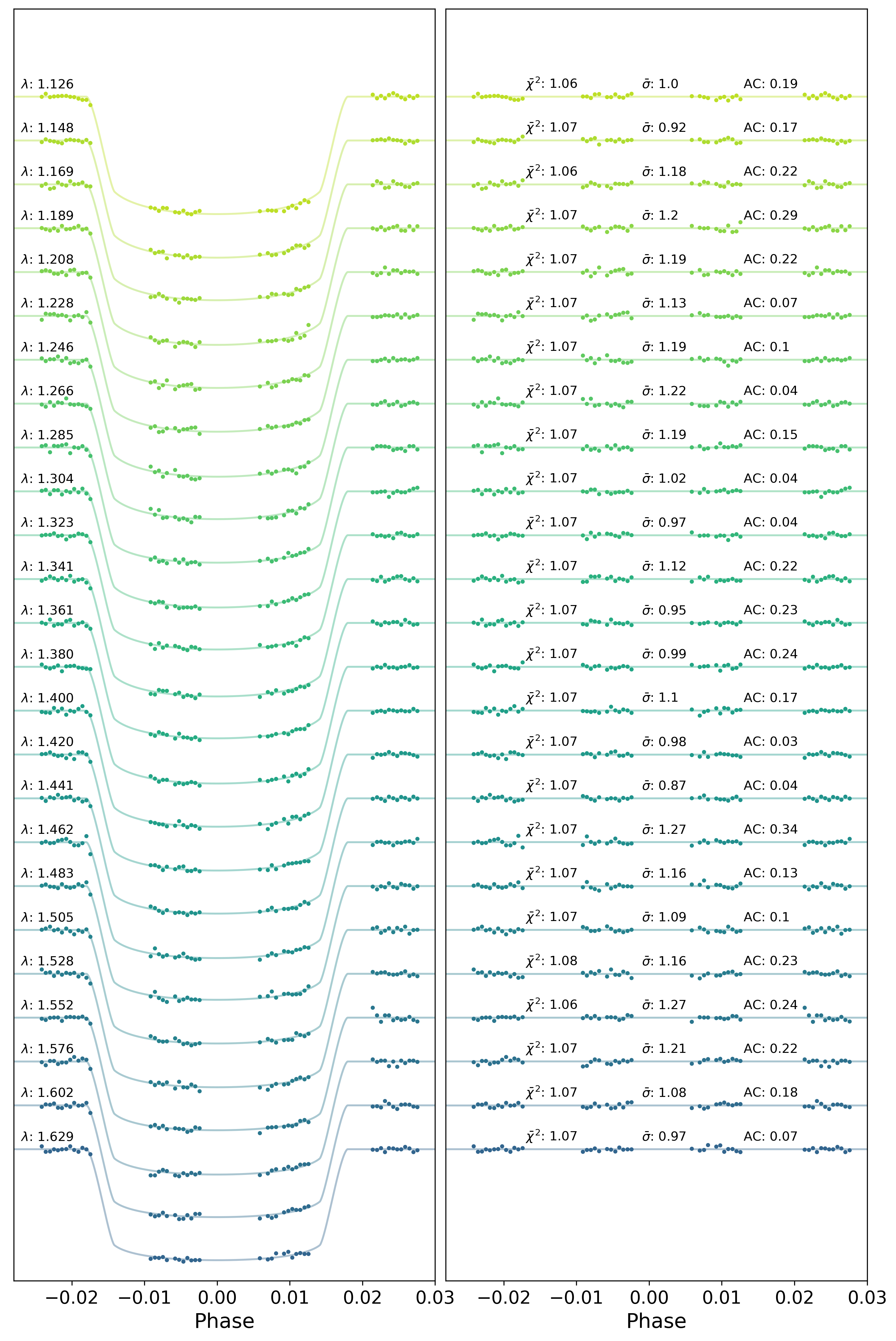}
\caption{Top figure: Results of the white light-curve of WASP-62\,b. Top: raw light-curve, after normalisation. Second: light-curve, divided by the best fit model for the systematics. Third: residuals. Bottom: auto-correlation function of the residuals. Bottom figure: Spectral light curves fitted with Iraclis for the transmission spectra where, for clarity, an offset has been applied. Left: the detrended spectral light curves with best-fit model plotted. Right: residuals from the fitting with values for the Chi-squared ($\chi^2$), the standard deviation of the residuals with respect to the photon noise ($\bar{\sigma}$) and the auto-correlation (AC).}
\label{fig:wasp62_lc}
\end{figure}

Our white light-curves were fit using literature values and the only free parameters, other than the coefficients for Hubble systematics, were the planet-to-star radius ratio and the transit mid time. This is motivated by the gaps in the observations, caused by Earth obscuration, which often means the ingress and egress of the transit is missed, limiting our ability to refine the semi-major axis to star radius ratio and the inclination planet's orbit. The limb-darkening coefficients were selected from the quadratic formula by \citet{Claret}, using the stellar parameters in Table \ref{tab:general_parameters}. Figure \ref{fig:white} shows the raw white light-curve, the detrended white light-curve, and the fitting residuals for WASP-127\,b while Figure \ref{fig:all} shows the fits of spectral light-curves for each wavelength bin. Similar plots for WASP-79\,b and WASP-62\,b are shown in Figures \ref{fig:wasp79_lc} and \ref{fig:wasp62_lc}.

\subsection{Atmospheric characterisation} \label{sec:atmochar}
The reduced spectra obtained using Iraclis were thereafter fitted using the publicly available\footnote{\url{https://github.com/ucl-exoplanets/TauREx3_public}} Bayesian atmospheric retrieval framework TauREx 3 \citep{al-refaie_taurex3}. 
TauREx uses the nested sampling code Multinest \citep{multinest} to explore the likelihood space of atmospheric parameters and features highly accurate line lists from the ExoMol project \citep{ExoMol}, along with those from HITRAN and HITEMP \citep{HITRAN,HITEMP}. In our retrieval analysis, we used 750 live points and an evidence tolerance of 0.5.
Several molecular opacities have been tested to model the spectra of the observations; in this publication, we considered five trace gases: H$_2$O \citep{polyansky_h2o}, CH$_4$ \citep{CH4}, CO\citep{li_co_2015}, CO$_2$ \citep{HITEMP}, NH$_3$ \citep{ExoMol_NH3} and FeH \citep{dulick_FeH}. In the wavelength range covered by G141, water vapour is the dominant spectral feature, but these other molecules can present detectable signals, particularly FeH \citep{tennyson_yurchenko}. Clouds are fitted assuming a grey opacity model.

\subsubsection{General setup}

In this study we use the plane-parallel approximation to model the atmospheres, with pressures ranging from $10^{-2}$ to $10^6$ Pa, uniformly sampled in log-space with 100 atmospheric layers. We included the Rayleigh scattering and the collision induced absorption (CIA) of H$_2$--H$_2$ and H$_2$--He \citep{abel_h2-h2, fletcher_h2-h2, abel_h2-he}. A summary of the fitted retrieval parameters is given in Table \ref{tab:general_parameters}. For consistency, the same parameter bounds have been applied for all three planets. Constant molecular abundance profiles were used, and allowed to vary freely between $10^{-12}$ and $10^{-1}$ in volume mixing ratio. The planetary radius was set to vary in our models between 0.5 $R_{ref}$ and 1.5 $R_{ref}$, where $R_{ref}$ the reference radius from the literature for each planet, as shown in Table \ref{tab:general_parameters}. This is assumed to be equivalent to the radius at $10^{6}$ Pa pressure.

The cloud top pressure ranged from $10^{-2}$ to $10^6$ Pa, in log-uniform scale. We consider a cloud top pressure of $10^6$ Pa to be a cloud-free atmosphere; the grey cloud model used for this study corresponds to a fully opaque layer below the cloud top pressure.

An isothermal atmosphere was assumed and the planetary temperature, $T_p$, set to vary from 400 to 2500 K; this is to accommodate the wide range in equilibrium temperatures between our three planets, which are between 1400 K and 1750 K as shown in Table \ref{tab:general_parameters}. 

\subsection{Atmospheric Detectability Index - ADI}
For quantifying the detection significance of an atmosphere, we use the Atmospheric Detection Index (ADI) from \citet{angelos30}, positively defined as the Bayes Factor between the nominal atmospheric model and the flat-line model (i.e. a model representing a fully cloudy atmosphere). For the flat line model, the only free parameters are the planet radius and temperature, along with the cloud pressure. The nominal model then includes Rayleigh scattering and the collision induced absorption of H$_2$--H$_2$ and H$_2$--He, as well as molecular opacities. If an atmosphere is detected at 3 $\sigma$ and 5 $\sigma$ level, the corresponding ADI will be above 3 and 11, respectively. An ADI below 3 suggests the atmospheric detection is not significant, indicating the spectral feature amplitudes are insufficient given the uncertainty of the data.

To quantify the detection of particular species, we computed the Bayes factor, which is the ratio of the Bayesian  evidences of different models. We follow the formalism by \cite{kass1995bayes} for model selection significance as well as 
translate the Bayes factor to the more traditional $\sigma$ significance nomenclature following \citep{BennekeSeager2012}.

\subsection{Ephemeris Refinement}

Accurate knowledge of exoplanet transit times is fundamental for atmospheric studies. To ensure the planets studied here can be observed in the future, we used our HST white light curve mid times, along with data from TESS \citep{ricker}, to update the ephemeris of each planet. TESS data is publicly available through the MAST archive and we used the pipeline from \cite{edwards_orbyts} to download, clean and fit the 2 minute cadence data. WASP-127\,b had been studied in Sector 9; WASP-79\,b in Sectors 4 and 5; and WASP-62\,b in Sectors 1-4 and 6-13. After excluding bad data, we recovered 4, 12 and 60 transits for WASP-127\,b, WASP-79\,b and WASP\,62b respectively. These were fitted individually with the planet-to-star radius ratio $R_p/R_s$, reduced semi-major axis ($a/R_s$), inclination ($i$) and transit mid time ($T_{mid}$) as free parameters. Finally, we fitted a linear period (P) to these mid times and selected the updated transit mid time (T$_0$) such that the co-variance between T$_0$ and P was minimised. Mid times were converted to BJD$_TDB$ using the tool from \citet{eastman}.

\section{Results}\label{sec:results}

Each planet's retrieval produced results consistent with the significant presence of water vapour, with opaque clouds in two of the three planets. While we did attempt to retrieve the carbon-based molecules, CO, CO$_2$, and CH$_4$, only their upper value could be constrained as they lack strong absorption features in the G141 wavelength range. In each case our best-fit solution also indicates the presence of FeH, with abundances of log(FeH) between -3.04 and -5.25.

\begin{figure}
    \centering
    \includegraphics[width = \columnwidth]{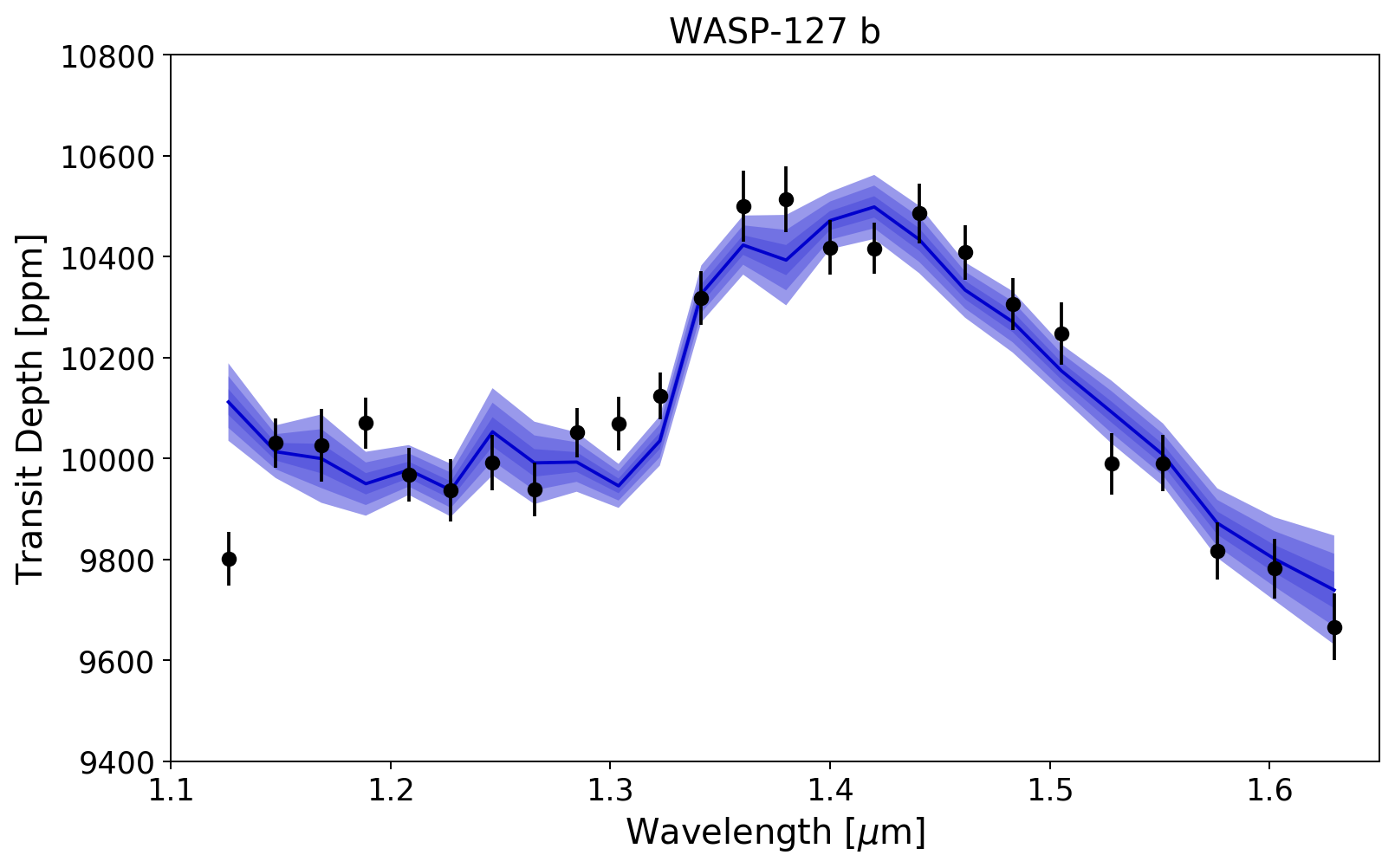}
    \includegraphics[width = \columnwidth]{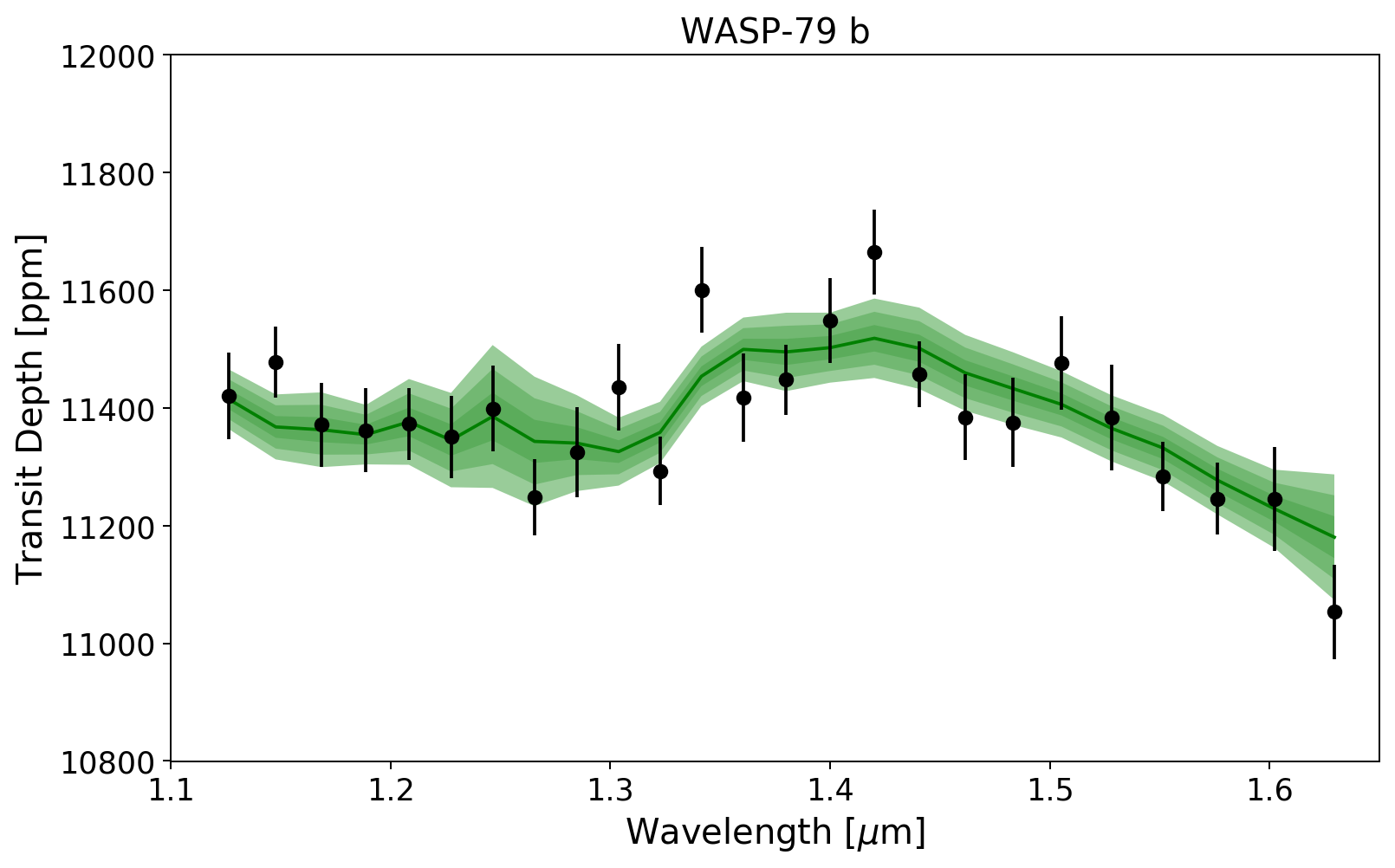}
    \includegraphics[width = \columnwidth]{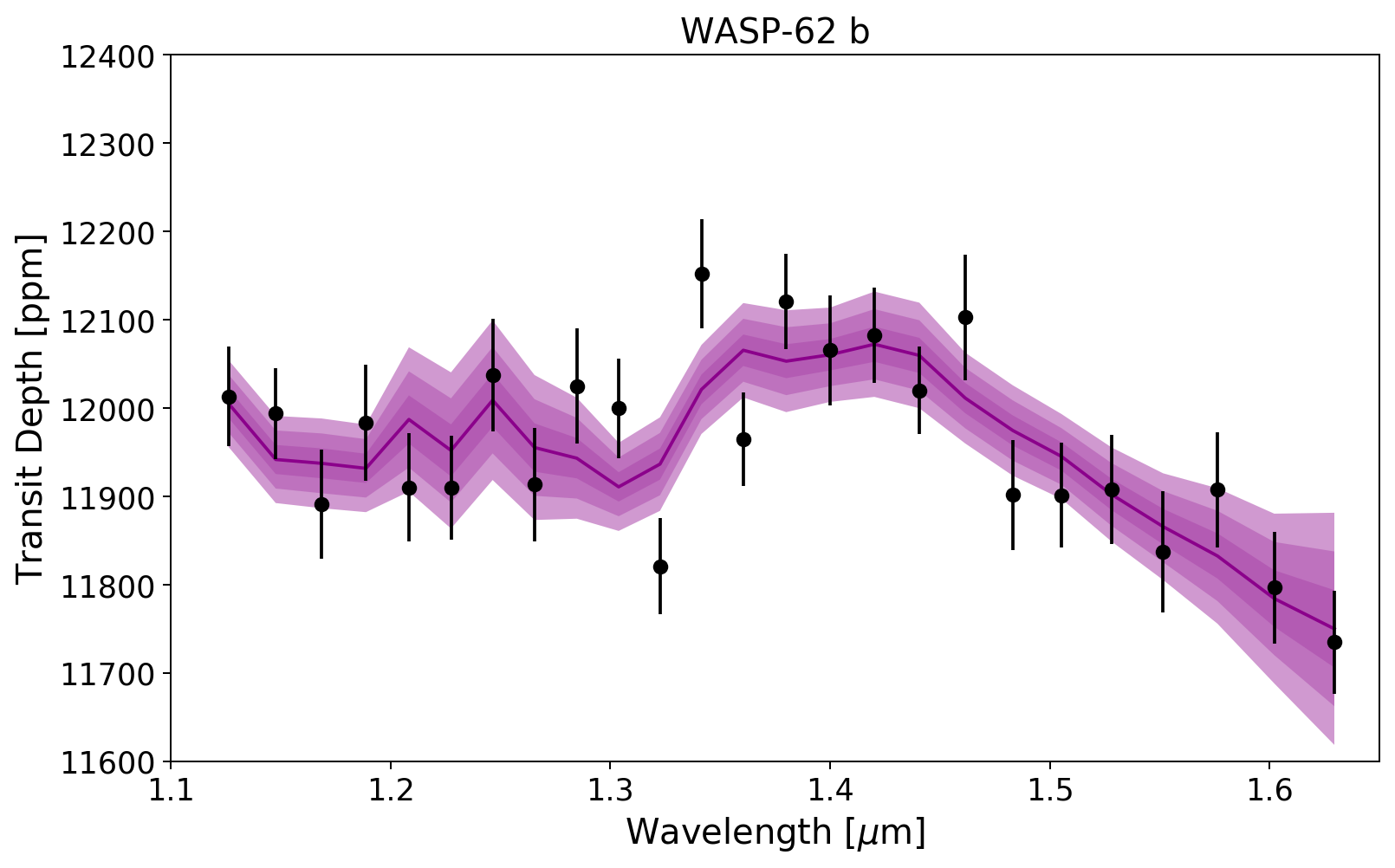}
    \caption{WFC3 spectral data and best-fit models, with 1-3$\sigma$ uncertainties, for the three planets: from top to bottom, WASP-127\,b; WASP-79\,b; and WASP-62\,b.}
    \label{fig:fit_spectra}
\end{figure}

The relatively high water abundances retrieved (10$^{-2}$-10$^{-3}$) for these three planets can be suggestive of metallicities in the super-solar regime \citep{solarmadhu,Pinhas_2018,Charnay_2018}. However, there are known degeneracies between the the cloud pressure, $10^6$\,Pa radius and water abundance retrieved from WFC3 data \citep{griffith,heng_10bar}. Additionally, due to the restrictions of the WFC3 wavelength ranges, these observations are not generally sensitive to the main carbon bearing species and arguments of high metallicities are usually based on retrieved water abundances alone, assuming that half of the oxygen is in H$_2$O as expected for a solar C/O ratio at high temperatures \citep{solarmadhu}. Hence, observations covering longer wavelength ranges are needed to further constrain the C/O ratios of these planets, to fully understand their metallicity.

\begin{figure}[ht!]
    \centering
    \includegraphics[width = \columnwidth]{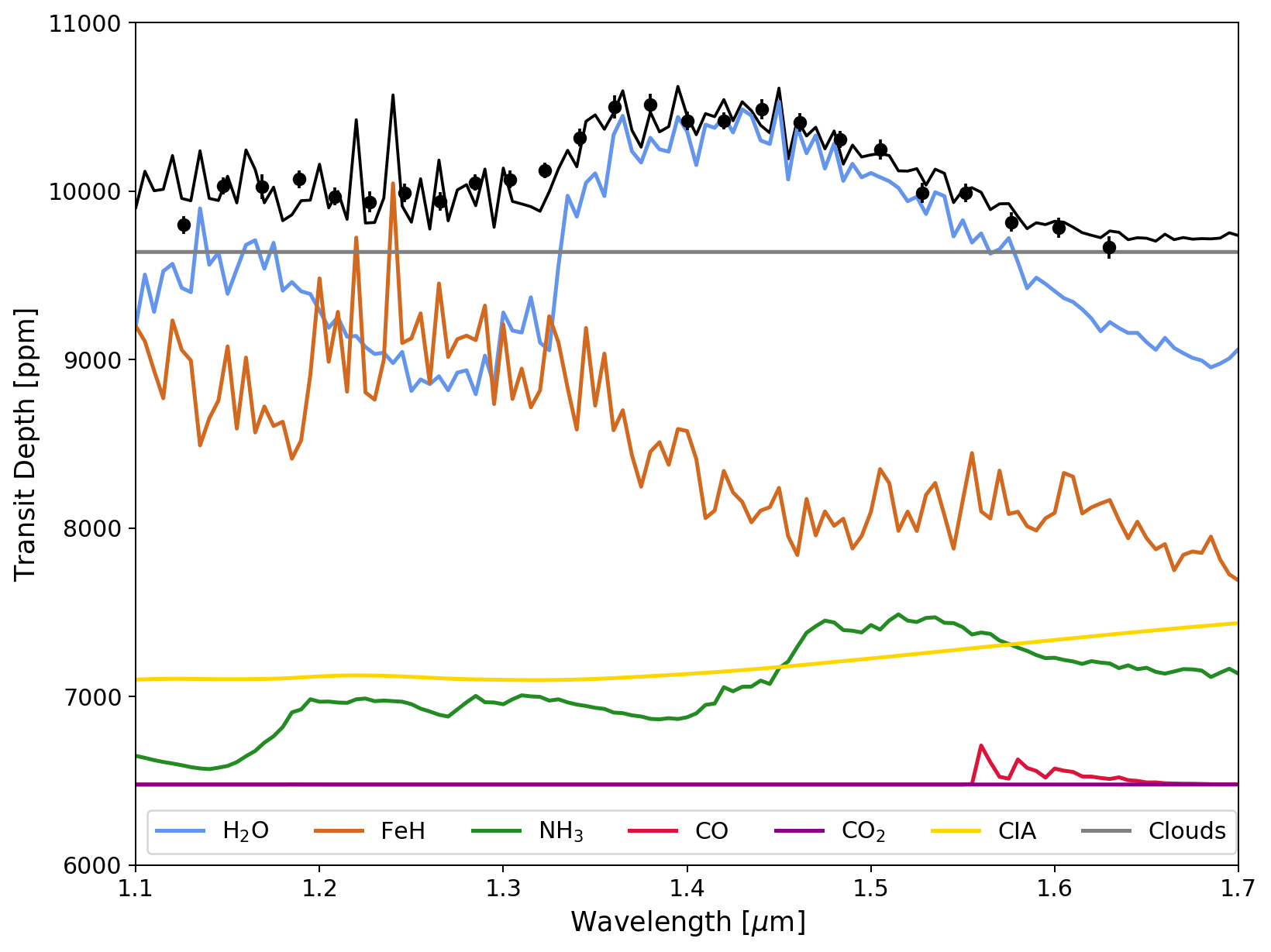}
    \includegraphics[width = \columnwidth]{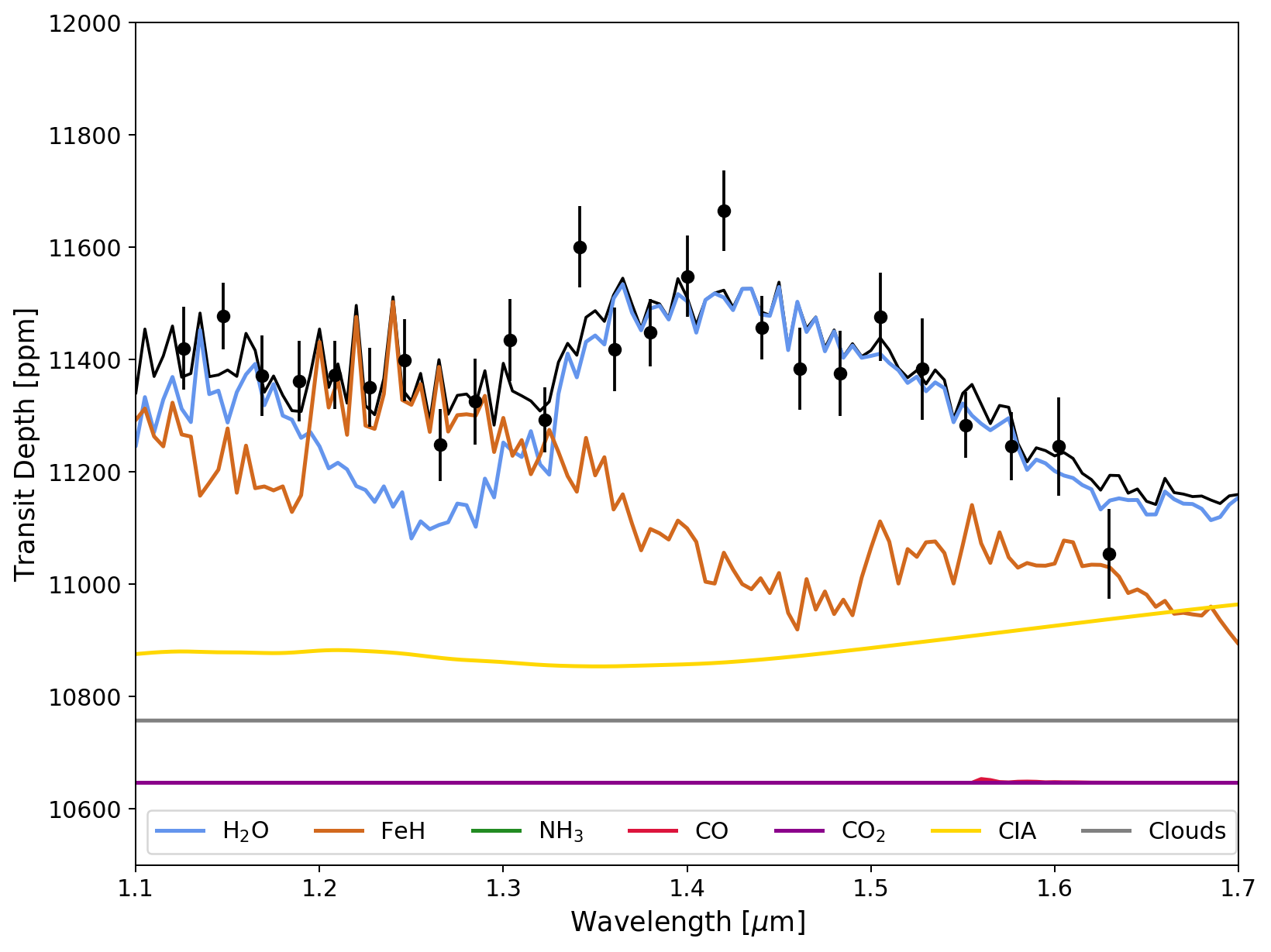}
    \includegraphics[width = \columnwidth]{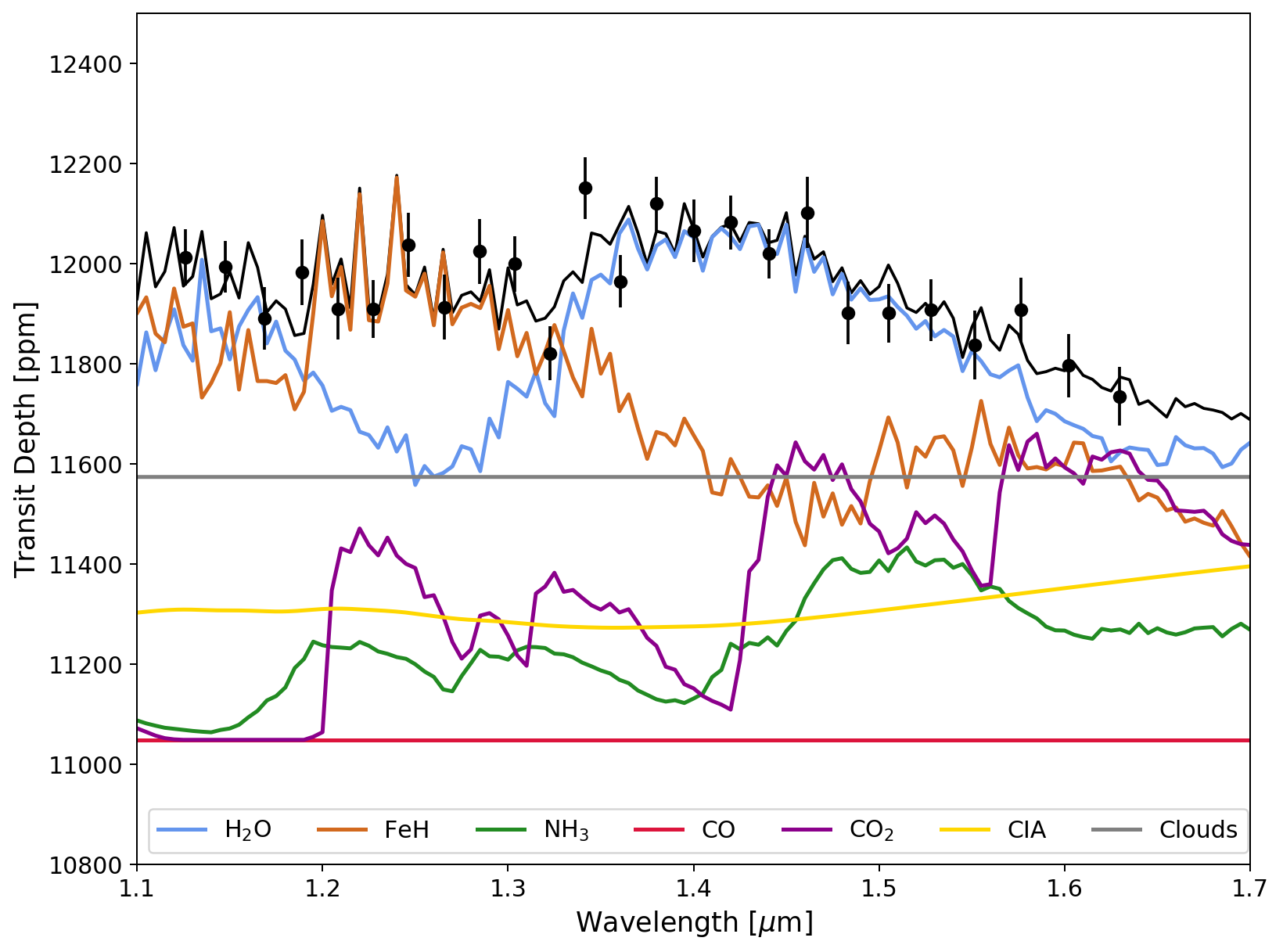}
    \caption{WFC3 spectral data and the contributions of active trace gases and clouds from the best-fit model for each planet. Top: WASP-127b. Middle: WASP-79b. Bottom: WASP-62b. In each case, the black line denotes the best-fit model.}
    \label{fig:fit_contrib}
\end{figure}

\begin{figure*}
    \centering
    \includegraphics[width=\textwidth]{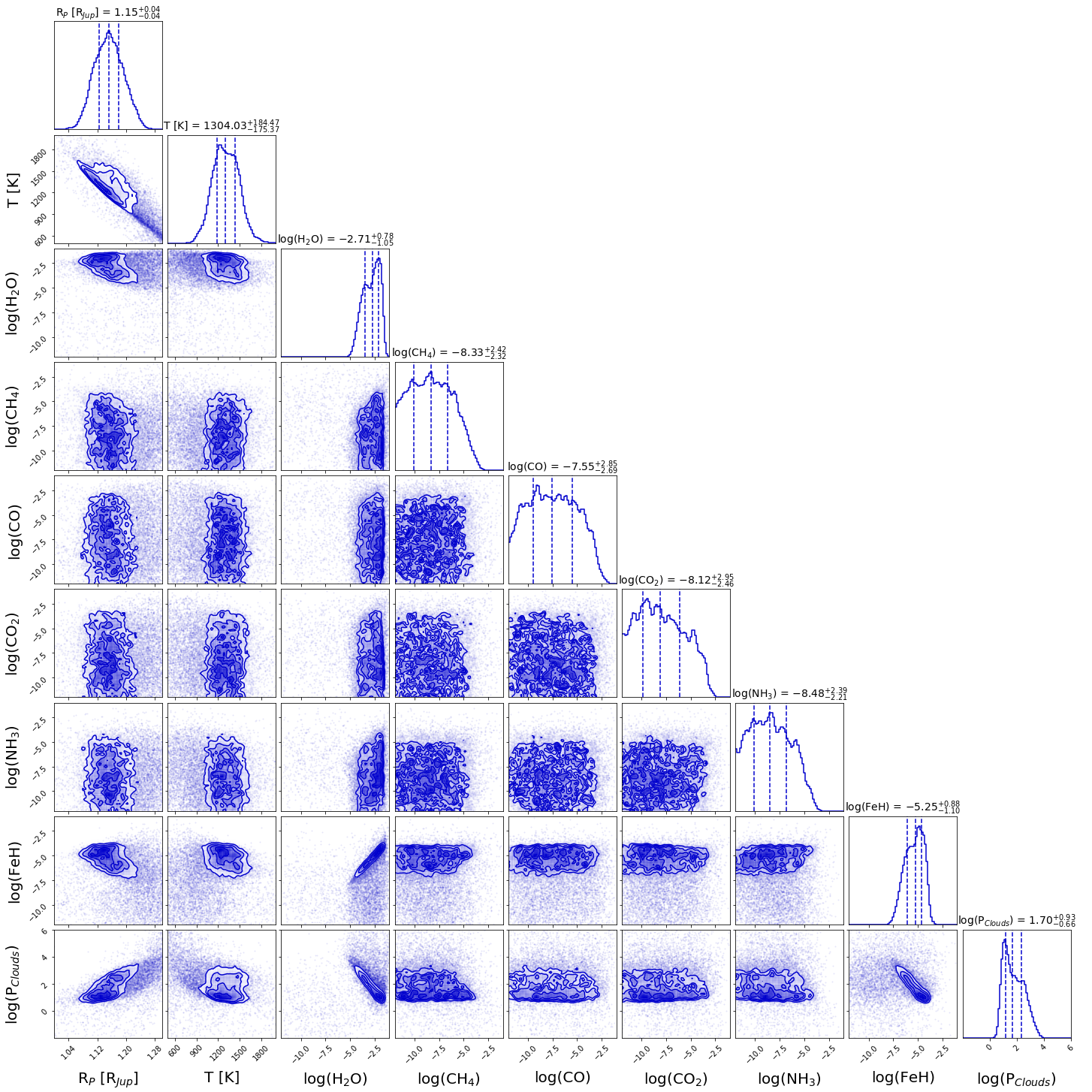}
    \caption{Posterior distributions from our WASP-127\,b retrieval.}
    \label{posteriors127}
    \centering
\end{figure*}

\begin{figure*}
    \centering
    \includegraphics[width=\textwidth]{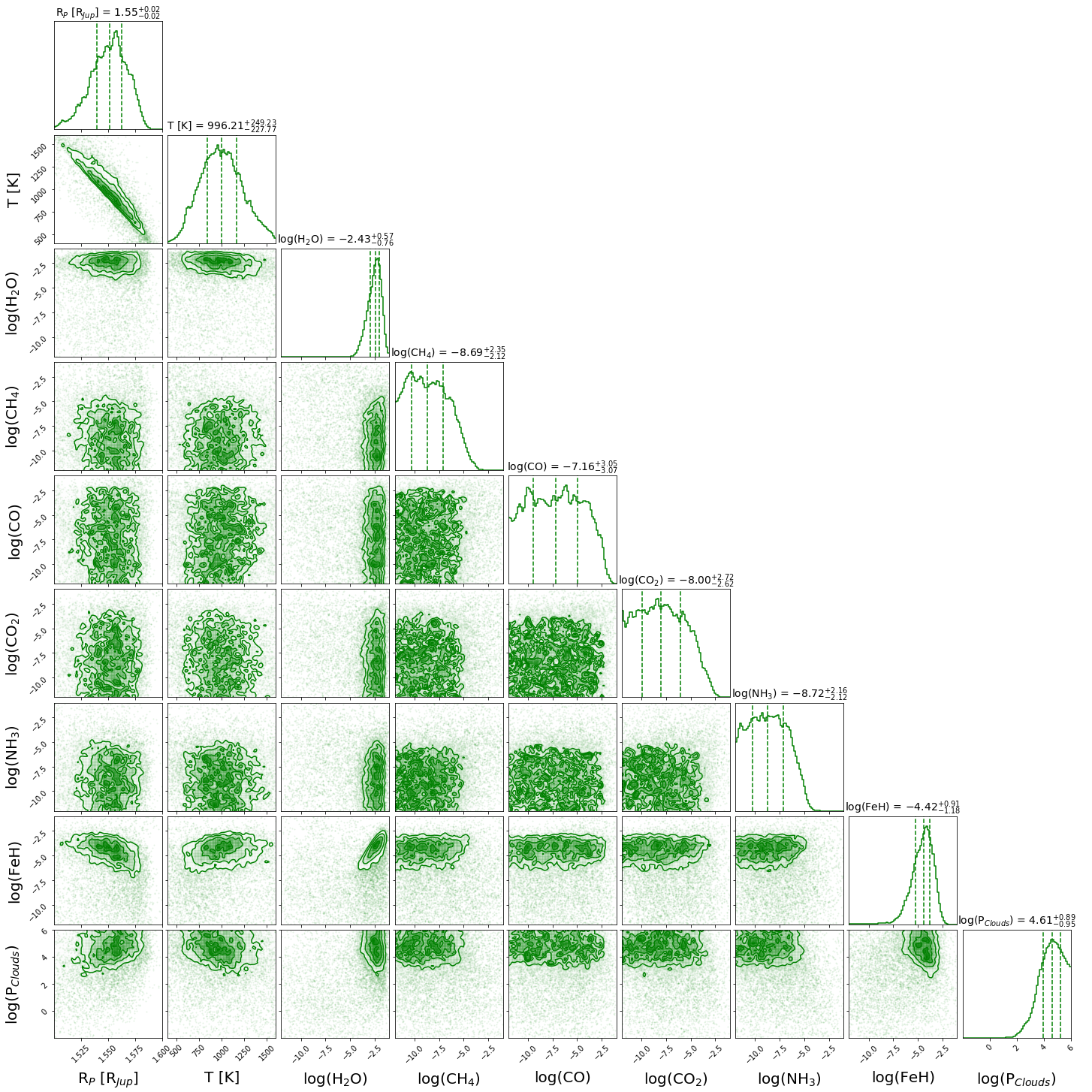}
    \caption{Posterior distributions for WASP-79\,b.}
    \label{fig:posteriors79}
\end{figure*}

Our findings are shown in Table \ref{tab:retrieval}, with Figure \ref{fig:fit_spectra} showcasing all three retrieved spectra with the corresponding contributions for each opacity source in Figure \ref{fig:fit_contrib}. For WASP-127\,b, the posteriors are shown in Figure \ref{posteriors127}, with equivalent results for WASP-79\,b in Figure \ref{fig:posteriors79} for WASP-62\,b in Figure \ref{fig:posteriors62}.\\


\begin{table*}
\centering
\caption{Table of fitted parameters for the retrievals performed on our targets}
\begin{tabular}{lllll}
\hline \hline
Retrieved Parameters  & Bounds & \textbf{WASP-127\,b} & \textbf{WASP-79\,b} & \textbf{WASP-62\,b} \\ \hline
$\log(H_2O)$  & 1e-12 - 1e-1  & $-2.71^{+0.78}_{-1.05}$ & $-2.43^{+0.57}_{-0.76}$ & $-2.03^{+0.52}_{-1.27}$ \\
$\log(FeH)$  & 1e-12 - 1e-1  & $-5.25^{+0.88}_{-1.10}$ & $-4.42^{+0.91}_{-1.18}$ & $-3.04^{+2.18}_{-2.27}$ \\
$\log(CH_4)$  & 1e-12 - 1e-1  & $< -5$ & $< -5$ & $< -5$ \\
$\log(CO)$  & 1e-12 - 1e-1  & $< -3$ & $< -3$ & $< -3$ \\
$\log(CO_2)$  & 1e-12 - 1e-1  & $< -3$ & $< -3$ & $< -3$ \\
$\log(NH3)$  & 1e-12 - 1e-1  & $< -5$ & $< -5$ & $< -5$ \\
$T_p$ [K] & 400-2500 & $ 1304^{+185}_{-175}$ & $996^{+249}_{-228}$ & $891^{+211}_{-164}$\\
$R_p\, [R_{J}]$ & $\pm $ 50\% & $1.15^{+0.04}_{-0.04}$ &  $1.55^{+0.02}_{-0.02}$ & $1.35^{+0.01}_{-0.02}$ \\ 
$\log(P_{clouds})$ & 1e-2 - 1e6 & $1.7^{+0.93}_{-0.66}$ & $> 4$ & $3.63^{+1.46}_{-1.29}$ \\
$\mu$ (derived) & & $ 2.34^{+0.20}_{-0.03}$ & $2.38^{+0.33}_{-0.07}$ & $2.39^{+0.51}_{-0.08}$ \\
\hline
ADI & - & 167.9 & 17.1 & 16.2  \\ 
\hline
$\sigma$-level & - & $>5\sigma$ & $>5\sigma$ & $>5\sigma$\\ \hline
\multicolumn{5}{c}{Updated Ephemeris}\\ \hline
P [days] & - & 4.1780619$\pm$1.3x10$^{-6}$ & 3.66239344$\pm$3.5x10$^{-7}$ & 4.41194014$\pm$7.4x10$^{-7}$ \\
T$_0$ [BJD$_{TDB}$-2450000] & - & 8238.943367$\pm$5.5x10$^{-5}$ & 8160.186968$\pm$3.9x10$^{-5}$ & 8476.084602$\pm$4.0x10$^{-5}$ \\
\hline \hline

\end{tabular}
\label{tab:retrieval}
\end{table*}


{}

\subsection{WASP-127\,b}
As expected given the low density, we retrieved a statistically significant atmosphere around WASP-127\,b with a strong detection of water and opaque clouds. The retrieved radius is 1.16$^{+0.04}_{-0.04}$ R$_J$ at a pressure of 10 bar, which is smaller than stated in previous studies (1.37 R$_J$, \citet{Chen}). However, our analysis is best fit with high altitude opaque clouds (log(P$_{clouds}$) = 1.85$^{+0.97}_{-0.66}$ Pa), which corresponds to approximately 1.37 R$_J$, thus explaining this difference between the retrieved radius and the radius in the literature.

In terms of chemistry, our best fit solution indicates significant amounts of water at log(H$_2$O)= -2.71$^{+0.78}_{-1.05}$, and constraints on FeH. FeH produces the flat absorption between 1.2 and 1.3 $\mu$m, whilst deepening the slope in the longer wavelengths (around 1.5 - 1.6 $\mu$m). We also note a correlation between the amount of these two molecules, the radius and the cloud pressure. For less H$_2$O and FeH, the model requires deeper clouds, but a higher base planet radius. 
In particular, the abundance of FeH can vary from $10^{-4}$ to $10^{-7}$, depending the complementary contribution of clouds. 

The posteriors for FeH are, however, always distinct; clouds cannot be used to completely replace the additional visible absorption provided by FeH. A lower metallicity, and larger radius, could be consistent with current data, but is not the best-fit solution. Given the posterior distribution, we don't find a clear correlation between the radius and the water abundance.

\subsection{WASP-79\,b}
For WASP-79\,b, we get very similar results to those of WASP-127\,b, with the exception of the cloud deck. Following our baseline approach, we find a large abundance of water at log(H$_2$O)= -2.43$^{+0.57}_{-0.76}$ and well defined constraints on the abundance of FeH with log(FeH)= -4.42$^{+0.91}_{-1.18}$. The clouds, however, do not impact the model, and we only retrieve a lower limit on their top pressure (P$_{clouds}$ \textgreater $10^3$ Pa). This means that either the planet possesses a clear atmosphere, or that the clouds are located below the visible pressure, at which the atmosphere is opaque due to molecular or collision induced absorption. We do not detect signatures of  CH$_4$, CO, or CO$_2$. 
The retrieved temperature of $\sim$1000 K is lower than the calculated equilibrium temperature for this type of planet; this was also found in \cite{sotzen_w79} and is discussed further in Section \ref{discussion}.

\subsection{WASP-62\,b}

The recovered spectrum of WASP-62\,b was flatter than the two other planets. However, we found that the data was best explained by the presence of H$_2$O and FeH and, for this retrieval, the recovered abundances are log(H$_2$O)=-2.03$^{+0.52}_{-1.27}$ and log(FeH)=-3.04$^{+2.18}_{-2.27}$. These results stem from detections in the lower-wavelength spectrum, below 1.5 $\mu$m, which guides the retrieval towards non-fully opaque sources, such as clouds and high-radius solutions. Again, the retrieved temperature is lower than the expected 1475 K equilibrium temperature, which is indicative of a large day-night temperature contrast and/or efficient cooling mechanisms. Our analysis indicates that clouds are likely to be present, but the quality of our data means that we cannot completely rule out a clear atmosphere.

The retrieved abundances are very high, but we note that the posteriors allow for a wide range of abundances and present interesting correlations; such as, the lower the abundance of H$_2$O and FeH are, the higher in the atmosphere the clouds are located. There is also a negative correlation between the molecular abundances and the temperature and, from the posterior distributions, we see that the data is consistent with abundances of order of $10^{-4}$ in H$_2$O and FeH.

Finally, we note that given the low spectral variations in this spectrum, the retrieval may lack a scale height constraint, which would provide a relevant baseline in predicting the molecular abundances and temperature more accurately.

\begin{figure*}
    \centering
    \includegraphics[width=\textwidth]{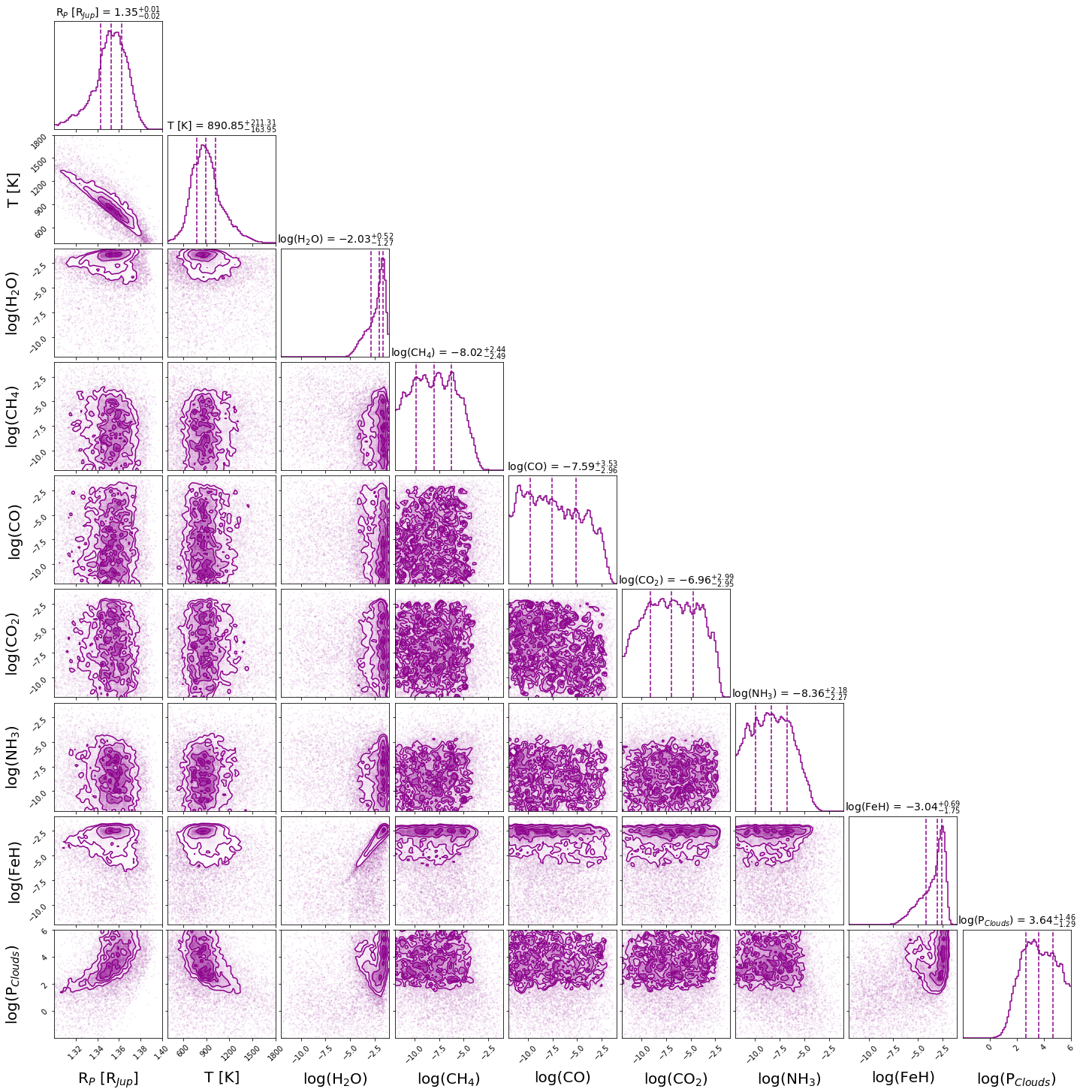}
    \caption{Posterior distributions for WASP-62\,b.}
    \label{fig:posteriors62}
\end{figure*}

\subsection{Ephemeris Refinement}

We found that the observed HST and TESS transits were consistent with literature ephemeris within 1$\sigma$. Nevertheless, we refined the period and reference mid transit time for each planet. The updated ephemeris is given in Table \ref{tab:retrieval} while the fitting for the TESS data can be seen in the figures of the appendix. The transit depth and errors of the WFC3 for the 3 planets are presented in Table \ref{tab:TransitdepthWFC3}.
The observed minus calculated plots are given in Figure \ref{fig:ephm_refine} and all transit mid times used for the fitting are listed in two Tables in the Appendix. These have been uploaded to ExoClock\footnote{\url{https://www.exoclock.space}}, an initiative to ensure transiting planets are regularly followed-up, keeping their ephemeris up-to-date for the ESA Ariel mission \citep{tinetti_ariel}.

\begin{figure}
    \centering
    \includegraphics[width=\columnwidth]{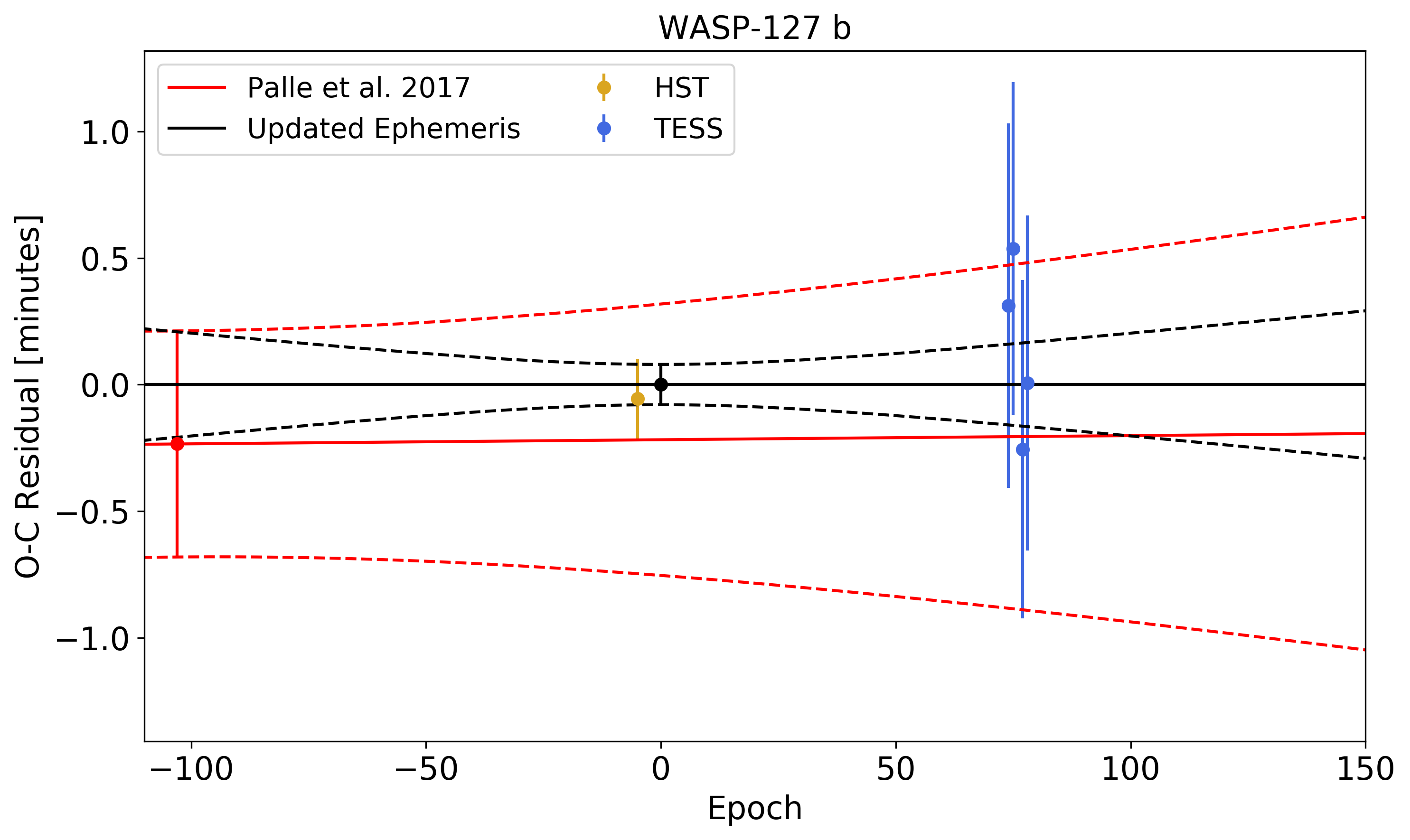}
    \includegraphics[width=\columnwidth]{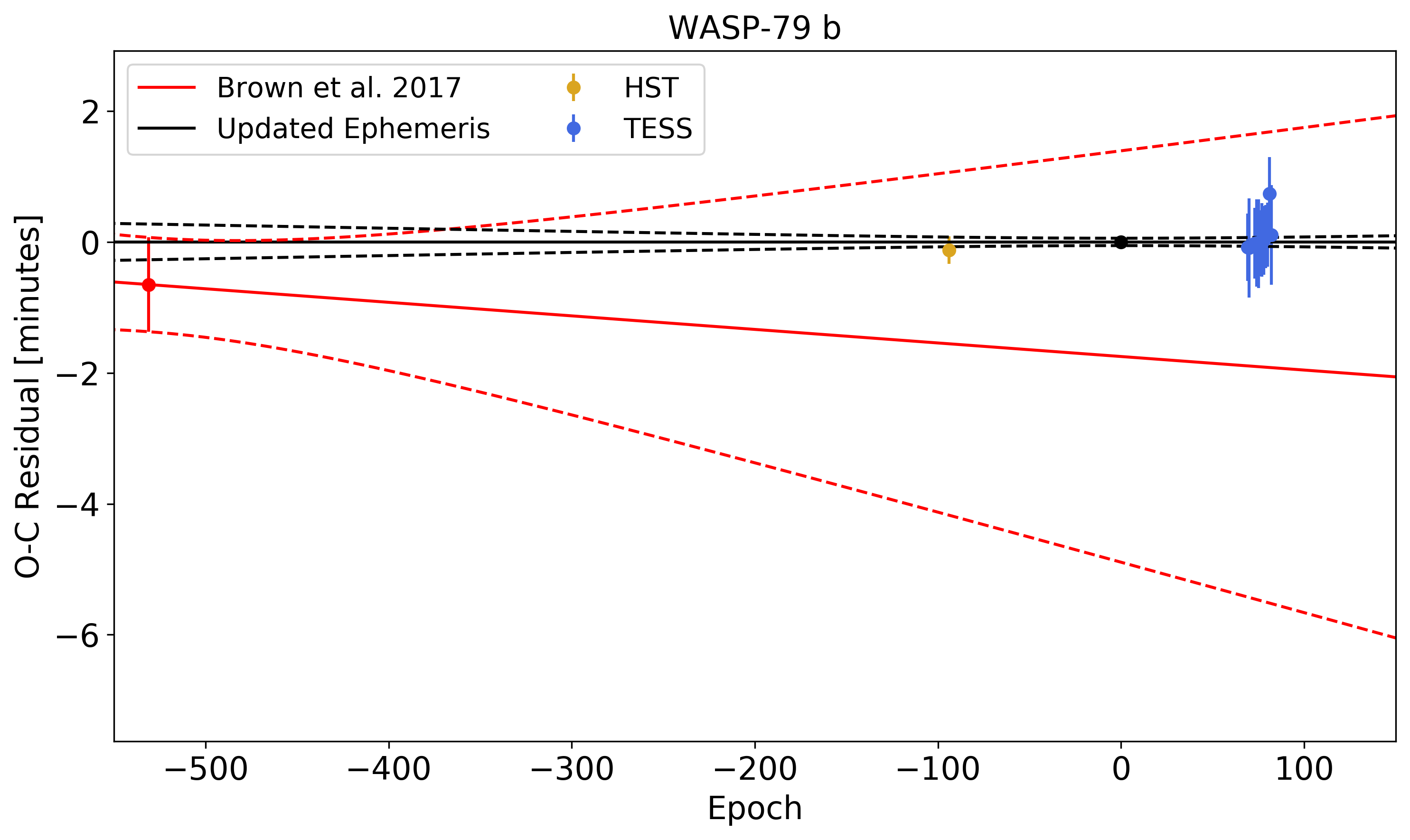}
    \includegraphics[width=\columnwidth]{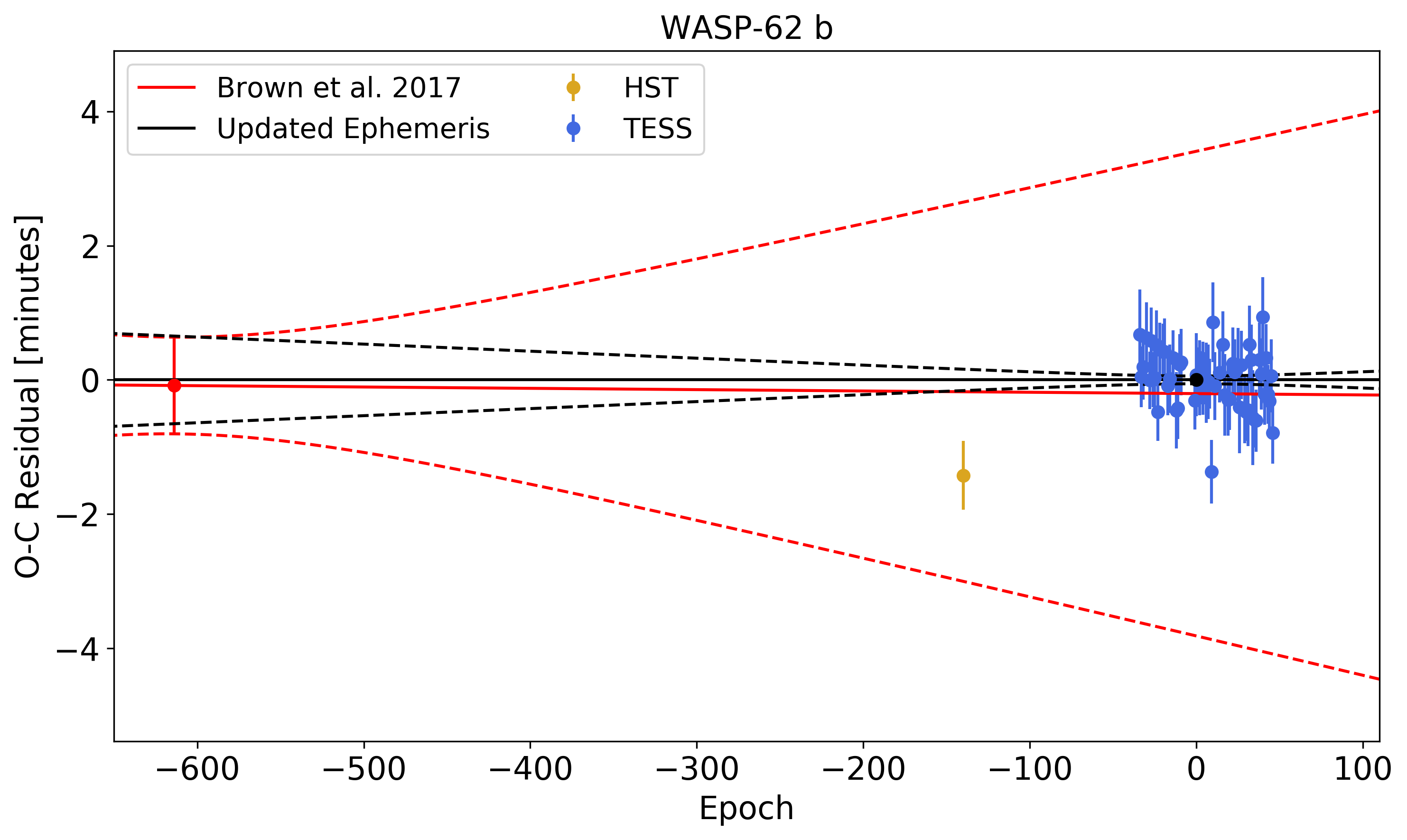}
    \caption{Observed minus calculated (O-C) transit mid times for WASP-127\,b (top), WASP-79\,b (middle) and WASP-62\,b (bottom). Transit mid time measurements from this work are shown in gold (HST) and blue (TESS), while literature T$_0$ values are in red. The black line denotes the new ephemeris of this work with the dashed lines showing the associated 1$\sigma$ uncertainties and the black data point indicating the updated T$_0$. For comparison, the previous literature ephemeris and their 1$\sigma$ uncertainties are given in red. In all cases, our results are compatible with the literature but provide a further refinement of the ephemeris.}
    \label{fig:ephm_refine}
\end{figure}{}


\begin{table}
    \centering
    \begin{tabular}{cccc}
    \hline
    Wavelength & WASP-127\,b & WASP-79\,b & WASP-62\,b\\\hline
1.12625 & 9800 $\pm$ 53 & 11420 $\pm$ 74 & 12012 $\pm$ 56 \\
1.14775 & 10030 $\pm$ 48 & 11477 $\pm$ 59 & 11993 $\pm$ 51 \\
1.16860 & 10026 $\pm$ 72 & 11371 $\pm$ 71 & 11890 $\pm$ 61 \\
1.18880 & 10070 $\pm$ 50 & 11362 $\pm$ 71 & 11983 $\pm$ 65 \\
1.20835 & 9967 $\pm$ 53 & 11372 $\pm$ 61 & 11909 $\pm$ 61 \\
1.22750 & 9936 $\pm$ 61 & 11350 $\pm$ 70 & 11909 $\pm$ 58 \\
1.24645 & 9990 $\pm$ 54 & 11398 $\pm$ 72 & 12036 $\pm$ 64 \\
1.26550 & 9938 $\pm$ 53 & 11248 $\pm$ 64 & 11913 $\pm$ 64 \\
1.28475 & 10050 $\pm$ 49 & 11325 $\pm$ 76 & 12024 $\pm$ 65 \\
1.30380 & 10068 $\pm$ 52 & 11434 $\pm$ 73 & 11999 $\pm$ 56 \\
1.32260 & 10123 $\pm$ 46 & 11292 $\pm$ 58 & 11820 $\pm$ 54 \\
1.34145 & 10318 $\pm$ 53 & 11600 $\pm$ 72 & 12151 $\pm$ 61 \\
1.36050 & 10500 $\pm$ 70 & 11417 $\pm$ 74 & 11964 $\pm$ 52 \\
1.38005 & 10513 $\pm$ 65 & 11448 $\pm$ 59 & 12120 $\pm$ 53 \\
1.40000 & 10417 $\pm$ 53 & 11548 $\pm$ 72 & 12065 $\pm$ 62 \\
1.42015 & 10415 $\pm$ 50 & 11664 $\pm$ 71 & 12082 $\pm$ 54 \\
1.44060 & 10485 $\pm$ 59 & 11457 $\pm$ 56 & 12019 $\pm$ 49 \\
1.46150 & 10408 $\pm$ 54 & 11384 $\pm$ 73 & 12102 $\pm$ 71 \\
1.48310 & 10305 $\pm$ 51 & 11375 $\pm$ 75 & 11901 $\pm$ 62 \\
1.50530 & 10246 $\pm$ 61 & 11476 $\pm$ 78 & 11901 $\pm$ 59 \\
1.52800 & 9989 $\pm$ 60 & 11383 $\pm$ 89 & 11907 $\pm$ 61 \\
1.55155 & 9990 $\pm$ 55 & 11283 $\pm$ 58 & 11837 $\pm$ 68 \\
1.57625 & 9816 $\pm$ 56 & 11245 $\pm$ 61 & 11907 $\pm$ 65 \\
1.60210 & 9781 $\pm$ 58 & 11245 $\pm$ 87 & 11796 $\pm$ 63 \\
1.62945 & 9665 $\pm$ 65 & 11053 $\pm$ 80 & 11734 $\pm$ 58 \\ \hline
    \end{tabular}

\caption{WFC3 transit depths and errors (in ppm) for for WASP-127\,b, WASP-79\,b and WASP-62\,b.}
\label{tab:TransitdepthWFC3}
\end{table}

\section{Discussion}\label{discussion}

Initially our baseline model did not include FeH, but these models struggled to fit the data, forcing solutions to lower temperatures and nonphysical values in order to account for the opacity sources at shorter wavelengths with a grey cloud deck. FeH has strong absorption features in the visible and near-infrared, and can be expected at the temperatures of these planets \citep{tennyson_yurchenko, madhu}; hence we propose it as the possible absorber to suit our spectral features and explore our justifications for FeH over molecules with similar spectral signatures, such as TiO or VO, in this discussion. FeH was not included in the analysis of \citet{angelos30} and thus, for the hotter planets in that study, retrievals with FeH may alter the retrieved atmospheric characteristics.

Theoretical equilibrium chemistry models predict FeH \citep{sharp,woitke} to be stable in the gas phase at the temperatures and pressures consistent with the planetary atmospheres considered here. FeH has previously been observed in L and M brown dwarfs at 1800 K  \citep{visscher_feh}. In cooler T dwarfs, it has been shown to appear where brown dwarfs have temperatures below 1350 K \citep{coolBD}, with some additional studies \citep{browndwarf} confirming FeH detection in dwarfs with temperatures of 1000 K. The latter of these detections is at temperatures comparable to the retrieved temperatures the planets here.

A recent study from \cite{iron_wasp76b} found atomic iron (Fe) in the day-side of the planet WASP-76\,b, and not in the terminator, concluding that Fe is condensing on the night-side, then falling into deeper layers of the atmosphere. Furthermore, the results in \cite{Pluriel} and \cite{Caldas2019} have investigated how the 3D structure of the atmosphere biases the abundances retrieved with typical retrieval codes, since there is a chemical dichotomy between the day- and night-side that is not considered in a 1D treatment of transit geometry.

We therefore identify three possible scenarios for the detection of FeH in these planets: \\

\begin{itemize}
    \item FeH is orginating from the day-side where the temperature is much higher, and leaks in the night-side before it is able to condense due to circulation processes \citep{HengCirculation}.
    \item Atmospheric retrieval studies involve temperature bias due to 3D effects, and we retrieve indeed a cooler temperature than expected; we discuss this in Section \ref{temperature}.
    \item A 3D effect is in play and we retrieve the FeH in the day-side inflated region of the limb \citep{Caldas2019,Pluriel}.\\
\end{itemize}

\begin{table*}
    \centering
    \caption{Comparison of the Bayesian log evidence for different models. For WASP-79\,b and WASP-62\,b, the retrieved temperature is always significantly below the equilibrium temperature for the planet, particularly if FeH is not included as an opacity source. In all cases, a better fit is obtained by including FeH.}
    \resizebox{\textwidth}{!}{
    \begin{tabular}{ccccc}\hline\hline
    \multicolumn{5}{c}{WASP-127\,b (No Molecules Log Evidence: 1.73  - Model (1))}\\\hline\hline
    Setup & Log Evidence & Sigma & Retrieved Temperature [K] & Equilibrium Temperature [K]\\\hline
    (2) H$_2$O, clouds &  161.87 & $>7$ w.r.t. (1) & 1027 & \multirow{4}{*}{1400$^\dagger$}\\
    (3) H$_2$O, CH$_4$, CO, CO$_2$, NH3, clouds & 161.27 & $<1$ w.r.t. (2) & 1005 &\\
    (4) H$_2$O, FeH, clouds & 170.20 & $4.48$ w.r.t. (2) & 1305 &\\
    (5) H$_2$O, CH$_4$, CO, CO$_2$, NH3, FeH, clouds & 169.65 & 4.49 w.r.t. (3) & 1304 &\\\hline\hline
    \multicolumn{5}{c}{WASP-79b (No Molecules Log Evidence: 173.33 - Model (1))}\\ \hline\hline
    Setup & Log Evidence & Sigma & Retrieved Temperature [K] & Equilibrium Temperature [K]\\\hline
    (2) H$_2$O, clouds &  187.77 & 5.72 w.r.t. (1) & 627 & \multirow{4}{*}{1716$^\ddagger$}\\
    (3) H$_2$O, CH$_4$, CO, CO$_2$, NH3, clouds & 187.88 & $<1$ w.r.t. (2) & 618 &\\
    (4) H$_2$O, FeH, clouds & 191.16 & 3.09 w.r.t. (2) & 948 &\\
    (5) H$_2$O, CH$_4$, CO, CO$_2$, NH3, FeH, clouds & 190.73 & 2.89 w.r.t. (3) & 996 &\\\hline\hline
    \multicolumn{5}{c}{WASP-62b (No Molecules Log Evidence: 176.35 - Model (1))}\\ \hline\hline
    Setup & Log Evidence & Sigma & Retrieved Temperature [K] & Equilibrium Temperature [K]\\\hline
    (2) H$_2$O,clouds &  187.22 & 5.03 w.r.t. (1) & 618 & \multirow{4}{*}{1475$^\ddagger$}\\
    (3) H$_2$O, CH$_4$, CO, CO$_2$, NH3, clouds & 186.66 & $<1$ w.r.t. (2) & 566 &\\
    (4) H$_2$O, FeH, clouds & 192.59 & 3.72 w.r.t. (2) & 890 \\
    (5) H$_2$O, CH$_4$, CO, CO$_2$, NH3, FeH, clouds & 192.31 & 3.80 w.r.t. (3) & 891 &\\\hline\hline
    \multicolumn{5}{c}{$^\dagger$ \cite{lam_wasp127}   $^\ddagger$ \cite{brown_rm}}\\\hline\hline
    \end{tabular}
    }
    \label{bayes_evid}
\end{table*}

Table \ref{bayes_evid} contains the log evidence of several retrievals for each planet. In all cases, the addition of FeH increases the goodness of fit, while also raising the retrieved temperature. We note that all models included clouds. By comparing the log evidence of the models with only H$_2$O and the models with H$_2$O and FeH, we confirm for all planets that clouds are not a suitable opacity substitute for FeH.

The difference in log evidence for these models $(\Delta\text{log}(E))$ is 8.33, 3.39, and 5.37 for WASP-127\,b, WASP-79\,b and WASP-62\,b respectively (4.48\,$\sigma$, 3.09\,$\sigma$ and 3.72\,$\sigma$ detection of FeH, respectively). This indicates strong to decisive evidence in favour of models containing FeH \citep{kass1995bayes, BennekeSeager2012, le2layer}.

While we postulate that our evidence holds for FeH, it is possible that we detect another, yet unidentified opacity source with absorption characteristics similar to those of FeH over the WFC3 passband. For instance, similar absorption features can be produced with metal oxides such as TiO, VO and YO. However, we do not expect the presence of these molecules in these planets due to the planets' low equilibrium temperatures. TiO and VO have condensation temperatures of over $\sim$2000K \citep{Lodders2002, HubenyTiOVO, FortneyTiOVO}; the highest equilibrium temperature featured of these three planets is WASP-79b's 1716K, as referenced in Table \ref{tab:general_parameters}, thus rendering it less likely that the spectral features are due to TiO and VO compared to FeH. Ultimately, this further exemplifies the need for longer wavelength coverage with JWST or Ariel to confirm the nature of observed absorption in the future.

For each planet, we have calculated the ADI and found significant evidence of atmospheric features for all three. Given the water detection on all three planets, our results support the conclusions drawn by \citet{angelos30}; inflated, hot Jupiter-like planets do not necessarily destroy water in their upper atmospheres.

\subsection{Retrieved temperature}
\label{temperature}

\begin{figure}
    \centering
    \includegraphics[width = \columnwidth]{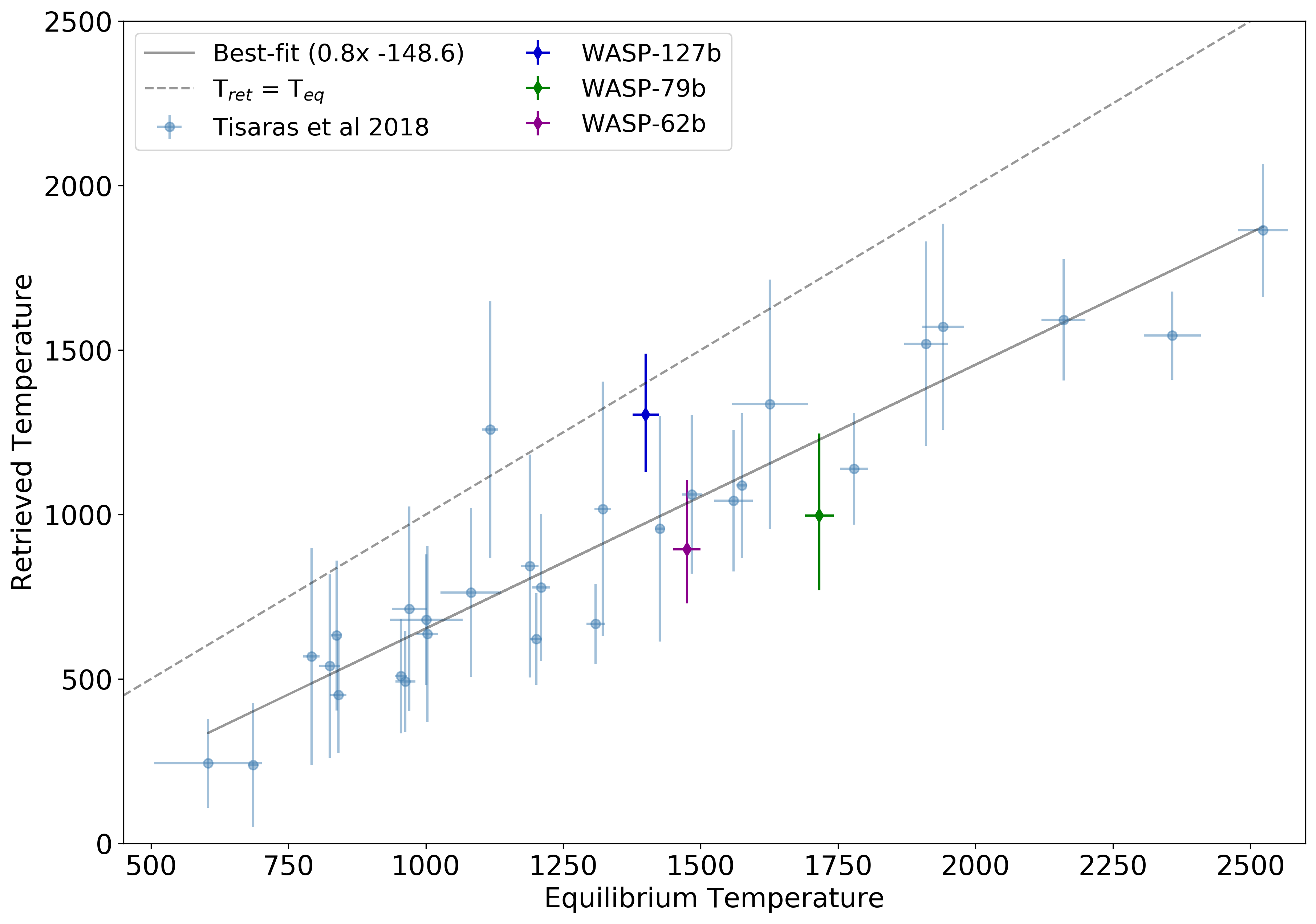}
    \caption{Correlation between the retrieved temperature and the equilibrium temperature for the planets studied in \cite{angelos30}. We observe a global trend that the retrieved temperature is lower than the calculated equilibrium temperature, and derived a best-fit for this trend. WASP-127b, WASP-79b and WASP-62b have been added; we can see that they follow this trend as well.}
    \label{fig:temperature}
\end{figure}

For the three planets considered, the temperature retrieved is notably lower than the equilibrium temperature. In Figure \ref{fig:temperature}, we present a plot analysing the temperatures retrieved for other planets, particularly giving attention to the population paper we based our study on, \cite{angelos30}. Indeed, retrieved temperatures are typically lower than the equilibrium ones, and we derived a best fit of this.

One of the key assumptions leading to this effect is that the equilibrium temperature is usually calculated for the planet day-side and considering a planetary albedo of zero. Considering an albedo greater than zero necessarily implies a loss of energy, and therefore a lower equilibrium temperature.

Furthermore, the region probed during transit eclipse observations is the terminator region: a mix between the day- and the night-side. The temperature difference observed may indicate a bias in the retrievals, which consider exclusively a 1D geometry of the atmospheres. This bias has been pointed out by several studies, especially \cite{Caldas2019,cold} and \cite{Pluriel}.

{}

\subsection{WASP-127\,b}

We used ExoREM (Exoplanet Radiative-convective Equilibrium Model; \cite{Baudino2017, Charnay_2018}), a self-consistent simulation software for brown dwarfs and giant exoplanets, to calculate the mean temperature profile and the expected abundances of WASP-127\,b assuming a solar composition.

The model suggests significant abundances of H$_2$O, CO and Na; though, as stated previously, the WFC3 coverage means our data set is only sensitive to H$_2$O.

\begin{figure}
\centering
\includegraphics[width=0.5\textwidth]{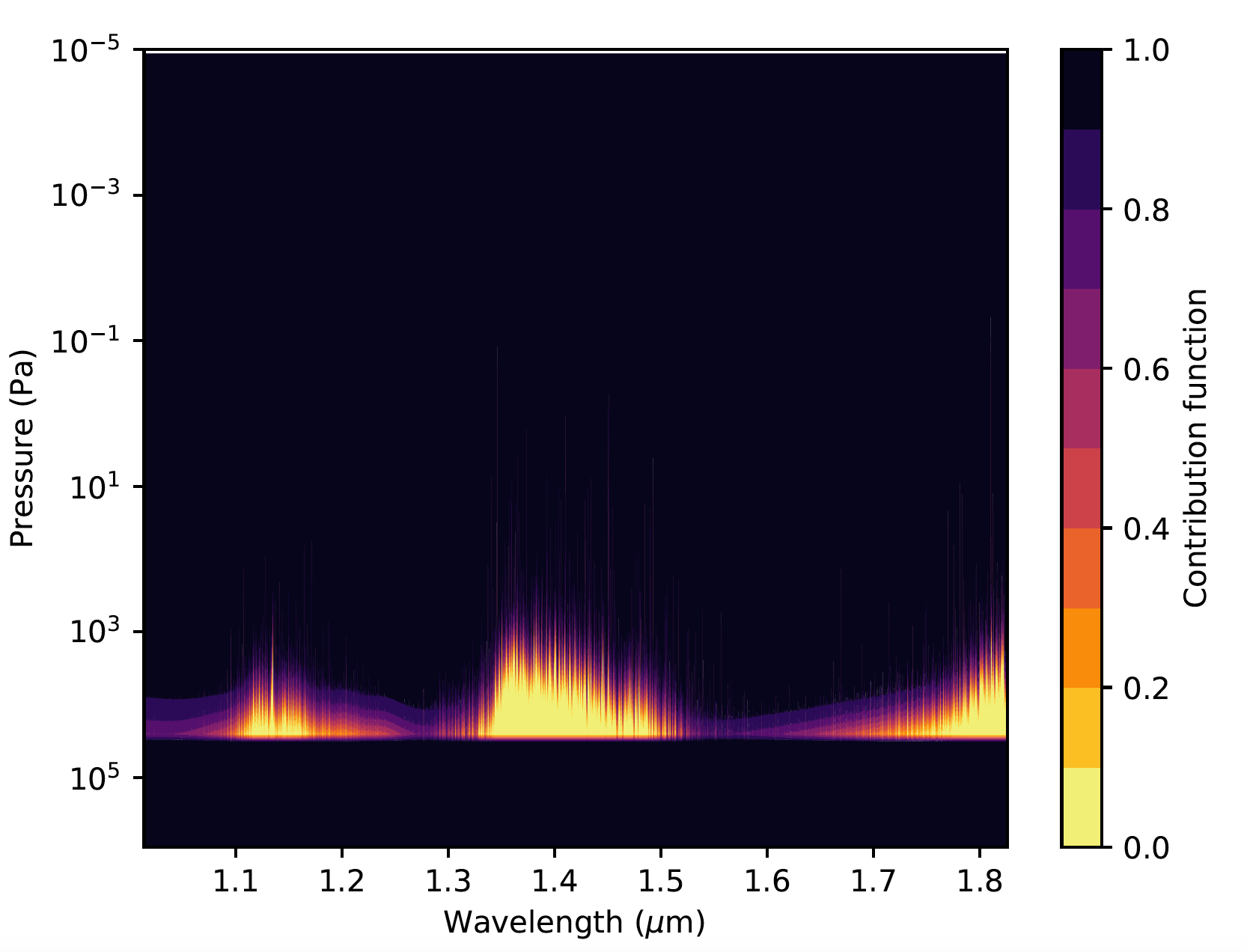}
\caption[Contribution function]{\label{fig:contribution_function} Typical contribution function of our retrievals. We can see the effect of the H$_2$O lines. The raise in sensitivity at $\approx$ 10$^4$ Pa is due to the opaque cloud.}
\end{figure}

\begin{figure}
\centering
\includegraphics[width=0.5\textwidth]{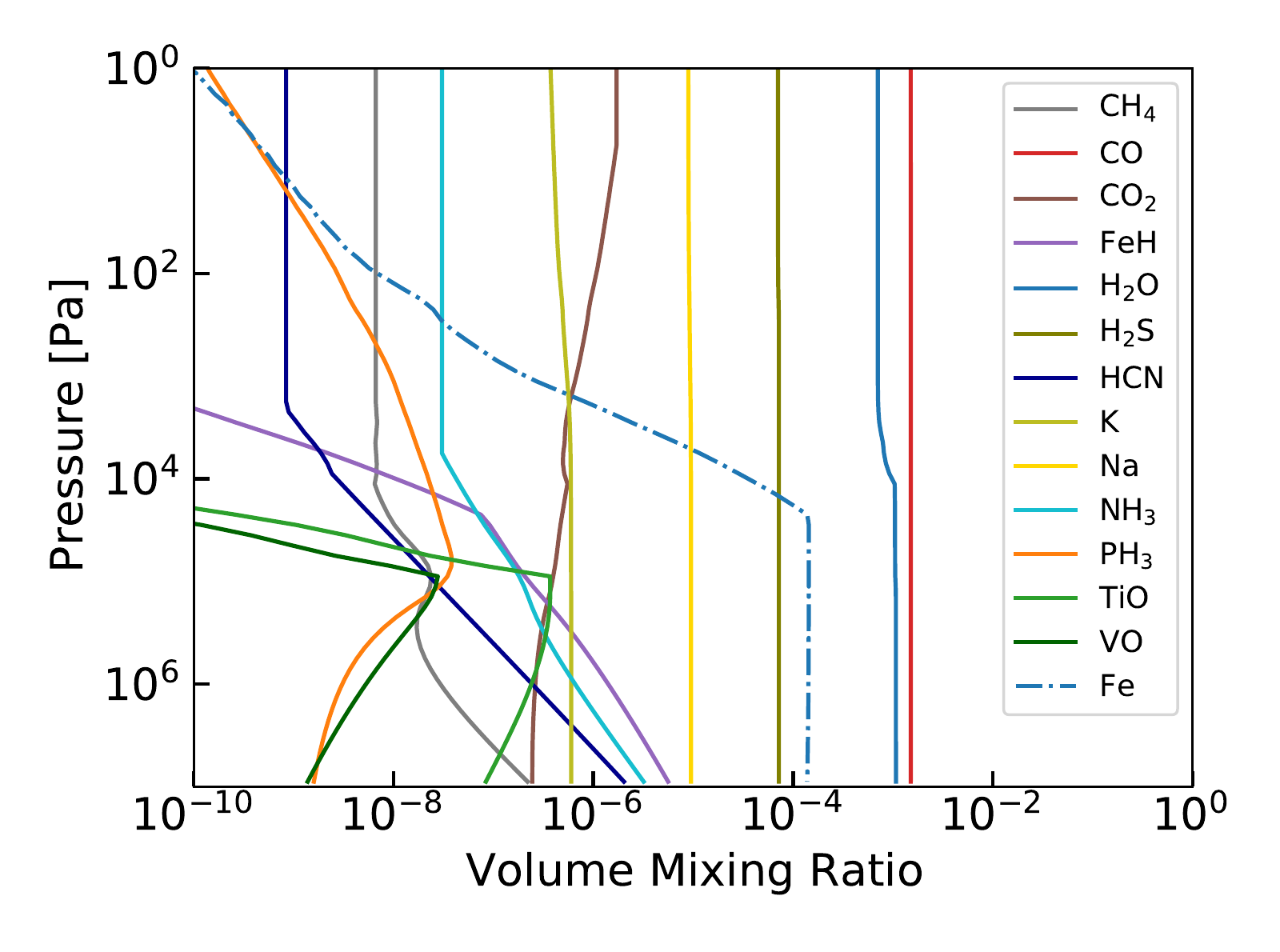}
\caption{Expected abundances of WASP-127\,b generated by Exo-REM, assuming chemical non-equilibrium for C, N and O bearing species, an eddy diffusion coefficient of 10$^8$ cm$^2$.s$^{-1}$, and a metallicity of 3 times solar \citep{Lodders2019}. We have included TiO and VO to the Exo-REM run but found their predicted abundances to be negligible ($<$ 10$^{-14}$).}

\label{fig:127b_abd}
\end{figure}

From Figure \ref{fig:contribution_function}, we can see that our retrieval is sensitive at pressures between $\approx 10^{4}$ and $10^{2}$ Pa. Figure \ref{fig:127b_abd} indicates the retrieved abundances of WASP-127\,b are compatible with a solar composition in this pressure range. The mean retrieved abundance of FeH is higher than expected values; however, the error spans three orders of magnitude, allowing for more physical solutions, as discussed at the beginning of this section.

\begin{figure}
\centering
\includegraphics[width=\columnwidth]{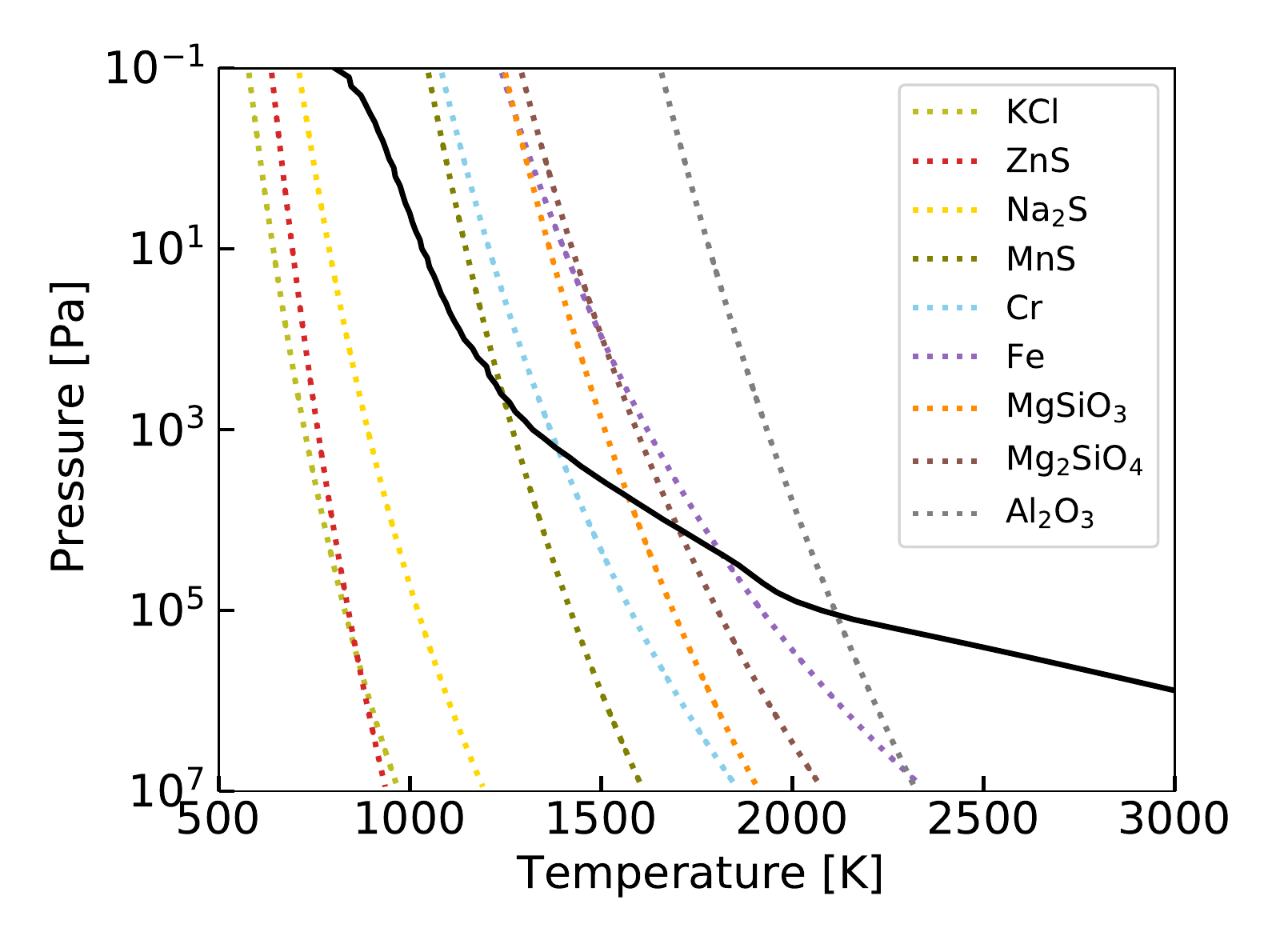}
\caption[WASP-127b temperature profile]{\label{fig:127b_TP} Mean temperature profile of WASP-127\,b, assuming radiative transfer equilibrium, generated by Exo-REM. The condensation profiles of various species are represented as dotted lines.}
\end{figure}

As displayed in Figure \ref{fig:127b_TP}, the retrieved temperature of WASP-127\,b is compatible with the calculated mean temperature profile within our pressure sensitivity range. We can also see that the calculated temperature profile crosses the condensation curves of MnS and Cr between $10^{4}$ and $10^{3}$ Pa. We could therefore expect clouds composed of these species to form at these pressures.

Low and high resolution spectra of WASP-127\,b have been collected with ground-based instruments. \citet{Palle} obtained low resolution spectroscopy with the Andalucia Faint Object Spectrograph and Camera (ALFOSC) spectrograph mounted on the Nordic Optical Telescope (NOT), covering the spectral range 0.45 - 0.85 $\mu m$. A slope was detected in the optical wavelength spectrum, interpreted as Rayleigh scattering and potentially Na. They also attribute the trend to TiO/VO with low significance. \citet{Chen} observed with the OSIRIS spectrograph, mounted on the Gran Telescopio Canarias (GTC) telescope, covering the range 0.4 - 1.0 $\mu m$. They presented detection of alkali metals (Na, K and Li), and hints of clouds and water with a retrieved abundance of log(H$_2$O)=-2.60$^{+0.94}_{-4.56}$. 
A recent study with Hubble STIS and WFC3, combined with Spitzer data from IRAC 1 and 2, also concluded that water was present in the atmosphere. \cite{spake2019} used an MCMC model to fit the data, resulting in a best-fit solution detecting H$_2$O, CO$_2$, Na and K; their water abundance of log(H$_2$O)=-2.87$^{+0.58}_{-0.61}$ is similar to the value retrieved here. 

Hence, our independent data reduction and analysis of the water content in the atmosphere of WASP-127\,b is consistent with these studies. We do not attempt a joint retrieval with this data due to the potential incompatibility between the data sets from different instruments as highlighted in \citet{yip}. Given that the planet lies within the short-period Neptunian desert and has large atmospheric features, it will be an intriguing target for further characterisation.

\subsection{WASP-79\,b}

\citet{sotzen_w79} utilised the same WFC3 data set, along with observations from ground-based facilities, TESS and Spitzer, to study the atmosphere of WASP-79\,b. Their retrieval results indicates the presence of H$_2$O, Na and FeH. Our retrieved water abundance is consistent with that from \citet{sotzen_w79}; (–2.20 \textless log(H$_2$O) \textless –1.55).

In our retrievals without FeH as an opacity source, our solution is driven to low temperatures; \cite{sotzen_w79} encountered a similar predicament when attempting to fit a chemical equilibrium model to the data. Here, by adding FeH as a retrieval parameter, our recovered temperature increases to 996 K$^{+249}_{-228}$, which more readily agrees with what is expected for the terminator region.

While the temperature is still cooler than expected, we note the degeneracy with the 10 bar radius. Our analysis of purely the HST/WFC3 data also favours the presence of H$_2$O and FeH. Na does not have features within the WFC3 spectral range, and we do not attempt the addition of other data for the aforementioned reasons.

\subsection{WASP-62\,b}

WASP-62\,b has demonstrably similar bulk characteristics to HD\,209458\,b; both planets have roughly the same radius and effective temperature, although HD\,209458\,b is $\approx$ 20\% more massive than WASP-62\,b \citep{bonomo2017}. Given their similarities, we may expect them to exhibit a similar atmospheric chemistry and structure. HD 209458b has been extensively analysed in the literature, with 3D simulations \citep{Showman2009} and cloud analysis \citep{Sing2016}, making it ideal to interpret the results on WASP-62\,b. We observe a cloud deck located at $\approx 2.5 \times 10^3$ Pa. This cloud deck could be explained by the condensation of MgSiO$_3$ in the atmosphere, as was the case with HD\,209458\,b \citep{Sing2016}.

Using the models of \citet{Showman2009} and \citet{Caldas2019}, we may expect the temperature at the terminator to be close to $\approx$ 1350 K. This is somewhat hotter than the $891^{+211}_{-164}$ K retrieved in our standard setup (2.2$\sigma$). However, as seen in the posterior plot in Figure \ref{fig:posteriors62}, there is a strong correlation between the temperature, the planet radius and the cloud pressure; so the data remains consistent with the expected temperature.

\subsection{Future Characterisation}

\begin{figure*}[ht]
    \centering
    \includegraphics[width = \columnwidth]{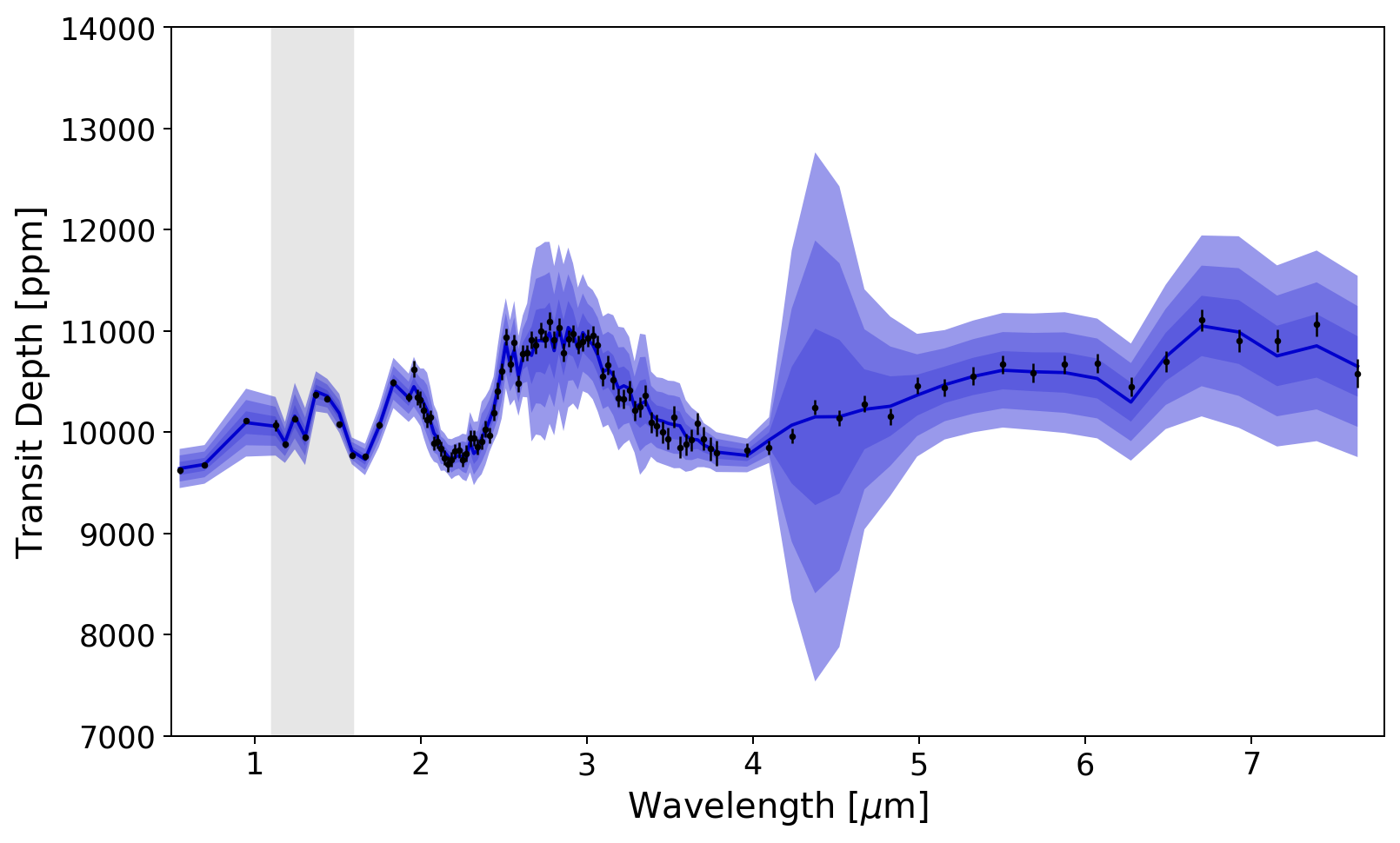}
    \includegraphics[width = \columnwidth]{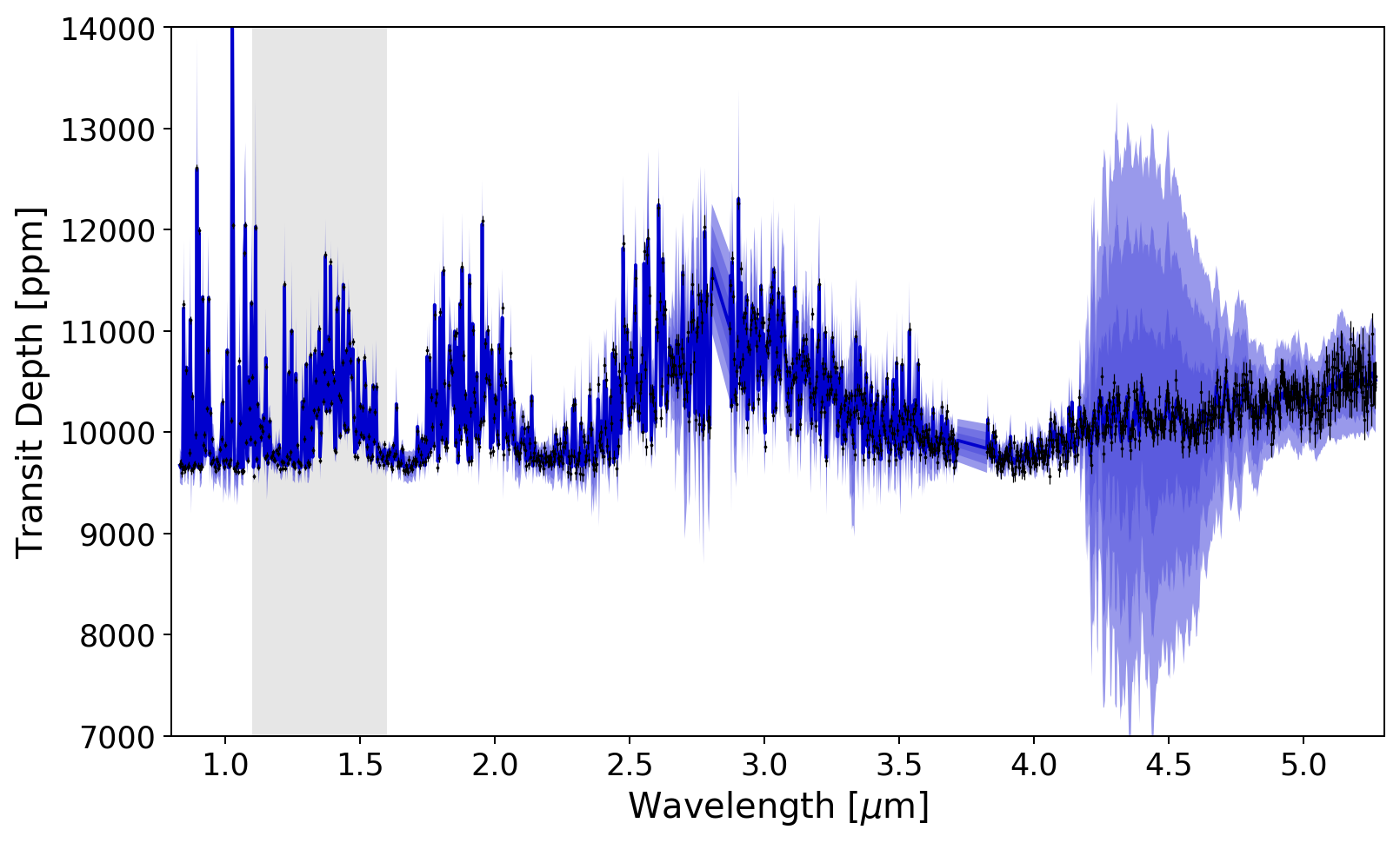}
    \includegraphics[width = \columnwidth]{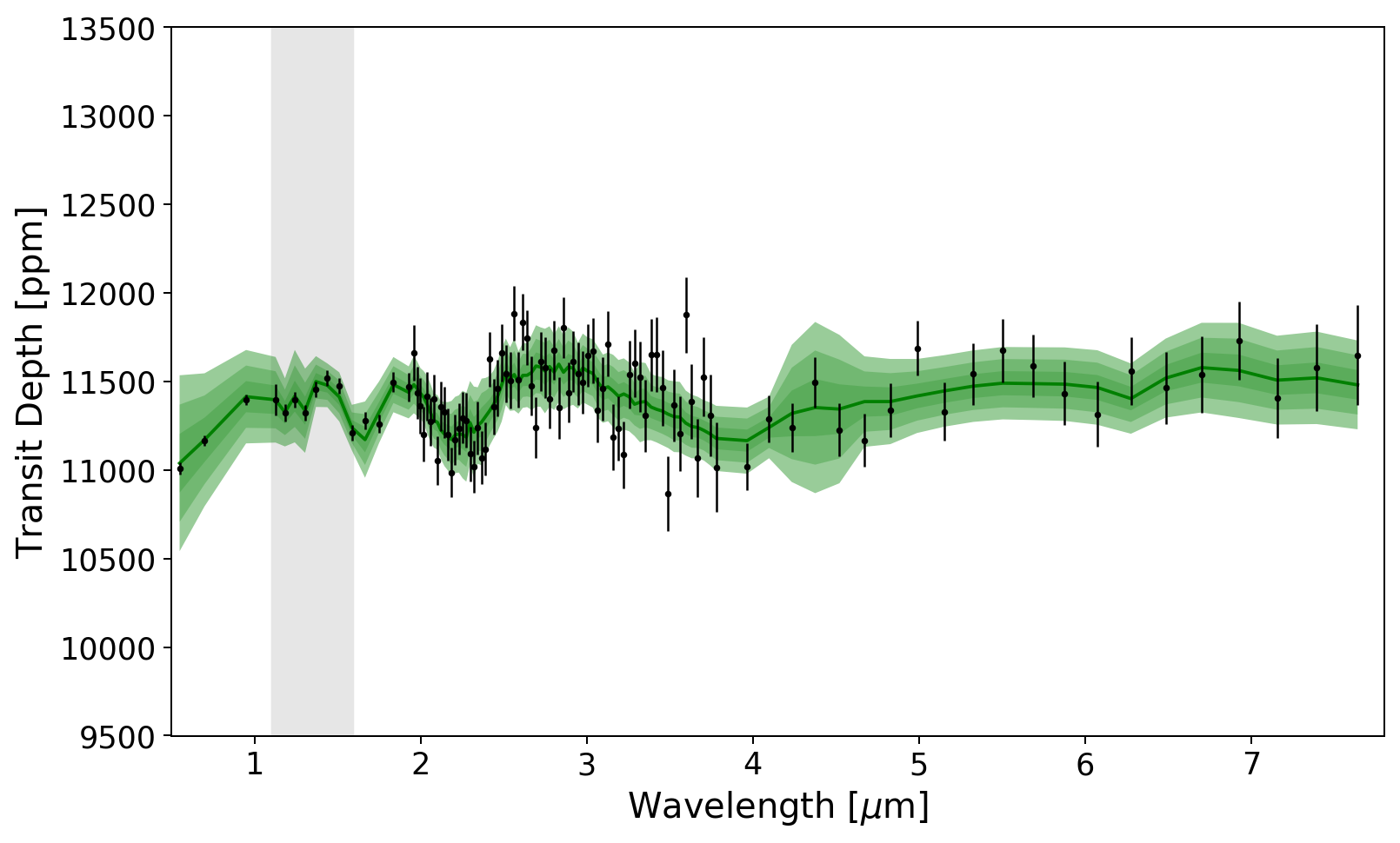}
    \includegraphics[width = \columnwidth]{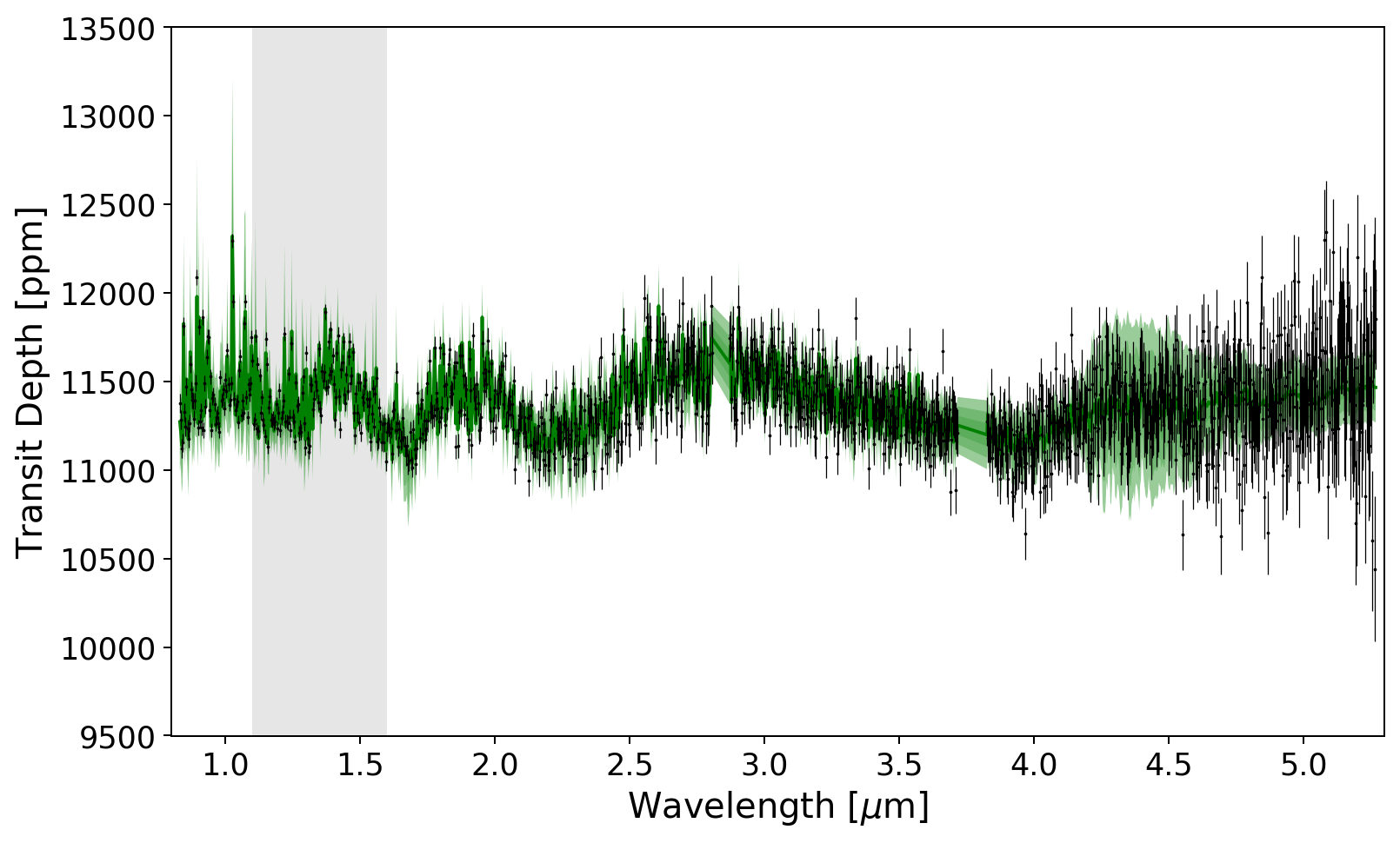}
    \includegraphics[width = \columnwidth]{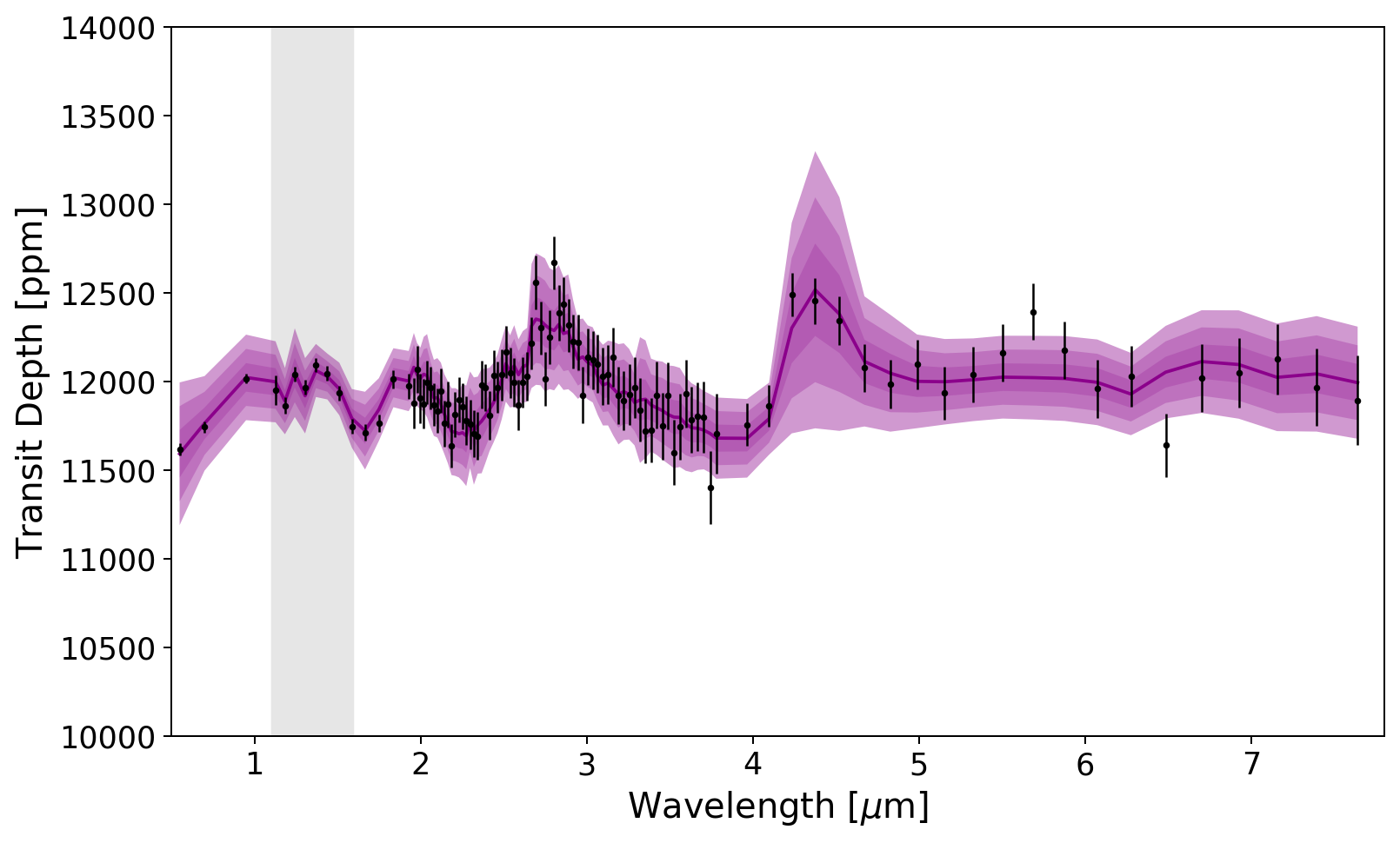}
    \includegraphics[width = \columnwidth]{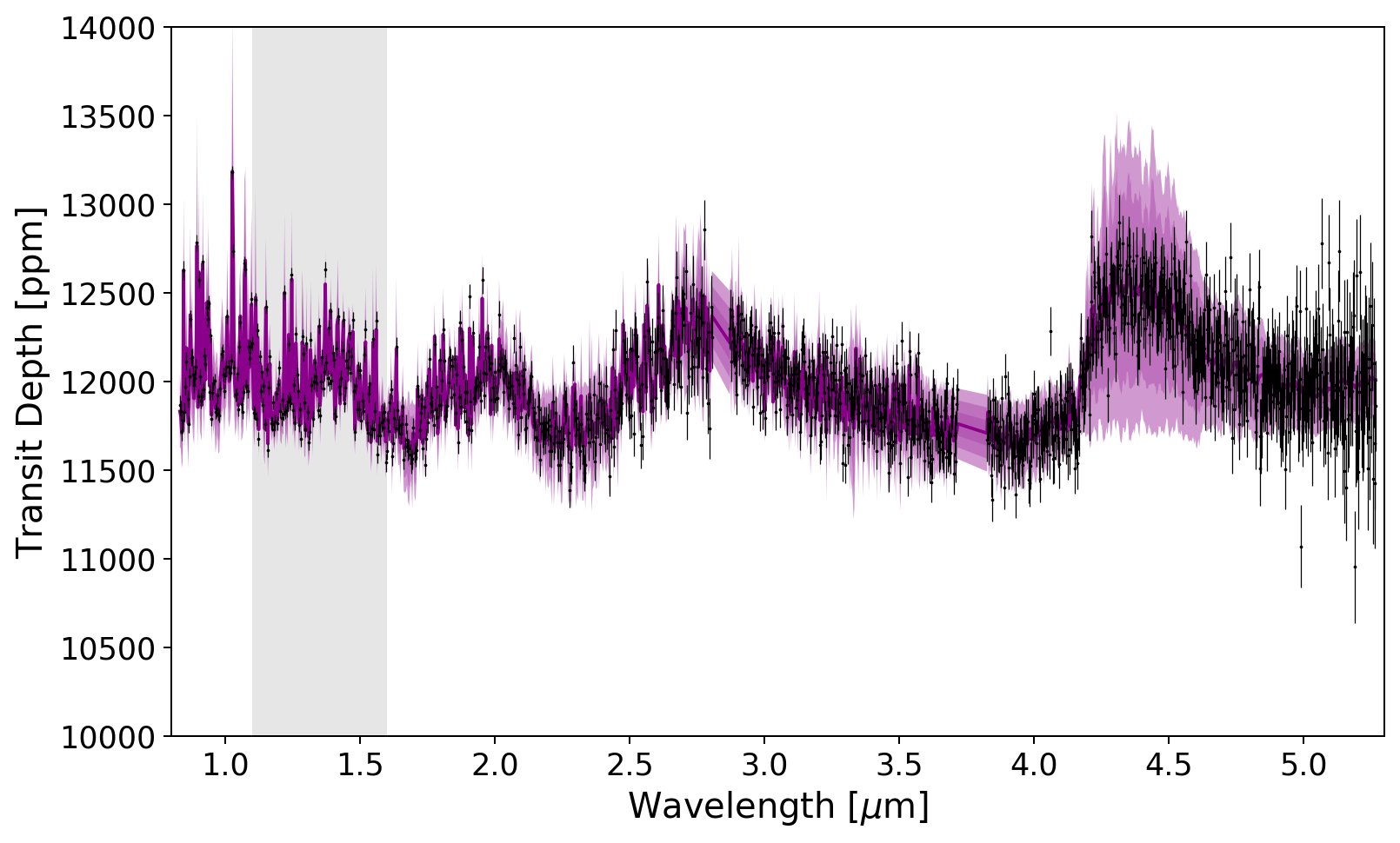}
    \caption{Simulated Ariel and JWST observations of the best-fit solutions retrieved in this work. For Ariel, WASP-127\,b is for a single observation while WASP-79\,b and WASP-62\,b have 3 stacked transits each. The transit depths have been offset to show the difference in the size of the atmospheric features between the planets. We note that, according to the work of \citet{GUILLY}, WASP-79\,b and WASP-62\,b may be more suited to study with Ariel via emission spectroscopy. JWST simulations have been performed using ExoWebb \citep{edwards_exowebb} for a single transit with NIRISS GR700XD as well as an observation with NIRSpec G395H.}
    \label{fig:ariel_plot}
\end{figure*}

Upcoming ground and space-based telescopes such as the European Extremely Large Telescope \citep[E-ELT,][]{brandl}, the Thirty Meter Telescope \citep[TMT,][]{skidmore}, the Giant Magellan Telescope \citep[GMT,][]{fanson}, the James Webb Space Telescope \citep[JWST,][]{greene}), Twinkle \citep{twinkle} and Ariel \citep{tinetti_ariel} will characterise the atmospheres of a large population of exoplanets via transit and eclipse spectroscopy at visible and infrared wavelengths. 
These missions will move the exoplanet field from an era of detection into one of characterisation, allowing for the identification of the molecular species present and their chemical profile, insights into the atmospheric temperature profile and the detection and characterisation of clouds \citep[e.g.][]{rocchetto,rodler,le2layer}.

Ariel has been selected as ESA's M4 mission adoption candidate for launch in 2028 and is designed for the characterisation of a large and diverse population of exoplanetary atmospheres to provide insights into planetary formation and evolution within our Galaxy. Ariel will provide simultaneous photometry and spectroscopy over 0.5 - 7.8 $\mu$m. Each of the planets studied here is an excellent target for atmospheric studies with Ariel \citep{GUILLY} and we use ArielRad \citep{mugnai} to simulate observations of this forthcoming mission. For each of the planets we take the best-fit solution from the Hubble WFC3 analysis to model Ariel observations at the native resolution of its instruments. Figure \ref{fig:ariel_plot} highlights the increased wavelength coverage and data quality that will be achieved with Ariel, allowing for a deeper understanding of each of these worlds. WASP-79\,b is part of the JWST  Early release Science (ERS) program and will be observed by JWST with several different instruments \citep{bean}. Here we simulate JWST observations for these planets, assuming NIRISS GR700XD and NIRSpec G395H are used. Again, the increase in data quality is easily discernible and, although it is not a dedicated exoplanet mission, JWST promises to provide exquisite data for atmospheric characterisation.

\section{Conclusion}

We have presented the analysis of data from Hubble's WFC3 G141 grism for three planets. By using the \verb+Iraclis+ pipeline and fitting the resultant spectra with TauREx, we have characterised the atmospheres of WASP-127\,b, WASP-79\,b and WASP-62\,b, recovering best fit models which favour the presence of H$_2$O and FeH in each case. This was performed during the ARES Summer School, using software and data publicly available to the community in order to allow for reproducible results. 

The properties of WASP-127\,b, particularly its extended atmosphere with clouds and large spectral features; the resultant high atmospheric detectability; and its unusually low density; make it an ideal target for further characterisation with the next generation of facilities. Large spectral features were also detected in WASP-79\,b and WASP-62\,b, with clouds in the atmosphere of the latter.

None of the three planets studied have strong features in their spectra that can be linked to NH$_3$, CH$_4$, CO, or CO$_2$. This is expected, given their spectroscopic lines do not have major bands in this wavelength range compared to the H$_2$O and FeH lines and higher quality data, with a broader spectral coverage, is required to improve constraints on the atmospheric chemistry. Nevertheless, studying the atmospheric composition of these planets has extended the catalogue of hot Jupiters studied with WFC3 from those by \cite{angelos30}. The ADI introduced therein has been utilised effectively in this paper to estimate the significance of these atmospheric observations. This was done in order to unify the statistical results between our study and that of further populations studies, which remain fundamental tools in understanding the nature and evolutionary history of planets.\\

\textbf{Acknowledgements:}\\

We want to thank the anonymous reviewer for the insightful and constructive comments which helped improve the quality of the manuscript.

This work was realised as part of ARES, the Ariel Retrieval Exoplanet School, in Biarritz in 2019. The school was organised by Jean-Philippe Beaulieu with the financial support of CNES. The publicly available observations presented here were taken as part of proposals 14619 and 14767, led by Jessica Spake and David Sing respectively. These were obtained from the Hubble Archive which is part of the Mikulski Archive for Space Telescopes. We are thankful to those who operate this archive, and public nature of which increases scientific productivity and accessibility \citep{Peek2019}.

JPB acknowledges the support of the University of Tasmania through the UTAS Foundation and the endowed Warren Chair in Astronomy, Rodolphe Cledassou, Pascale Danto and Michel Viso (CNES). BE, QC, MM, AT and IW rfunding from the European Research Council (ERC) under the European Union's Horizon 2020 research and innovation programme grant ExoAI (GA No. 758892) and the STFC grants ST/P000282/1, ST/P002153/1, ST/S002634/1 and ST/T001836/1. NS acknowledges the support of the IRIS-OCAV, PSL. MP acknowledges support by the European Research Council under Grant Agreement ATMO 757858 and by the CNES. RB is a Ph.D. fellow of the Research Foundation--Flanders (FWO). WP, TZ, and AYJ have received funding from the European Research Council (ERC) under the European Union's Horizon 2020 research and innovation programme (grant agreement n$^\circ$ 679030/WHIPLASH). OV thank the CNRS/INSU Programme National de Plan\'etologie (PNP) and CNES for funding support. GG acknowledges the financial support of the 2017 PhD fellowship programme of INAF. LVM and DMG acknowledge the financial support of the ARIEL ASI grant n. 2018-22-HH.0.

This paper includes data collected by the TESS mission which is funded by the NASA Explorer Program. TESS data is publicly available via the Mikulski Archive for Space Telescopes (MAST). \\

\vspace{3mm}
\textbf{Software:} Iraclis \citep{Iraclis}, TauREx3 \citep{al-refaie_taurex3}, pylightcurve \citep{tsiaras_plc}, ExoTETHyS \citep{morello_exotethys}, ArielRad \citep{mugnai}, ExoWebb \citep{edwards_exowebb}, Astropy \citep{astropy}, h5py \citep{hdf5_collette}, emcee \citep{emcee}, Matplotlib \citep{Hunter_matplotlib}, Multinest \citep{multinest}, Pandas \citep{mckinney_pandas}, Numpy \citep{oliphant_numpy}, SciPy \citep{scipy}.\\

\textbf{Appendix:}\\

\begin{figure}
    \centering
    \includegraphics[width=\columnwidth]{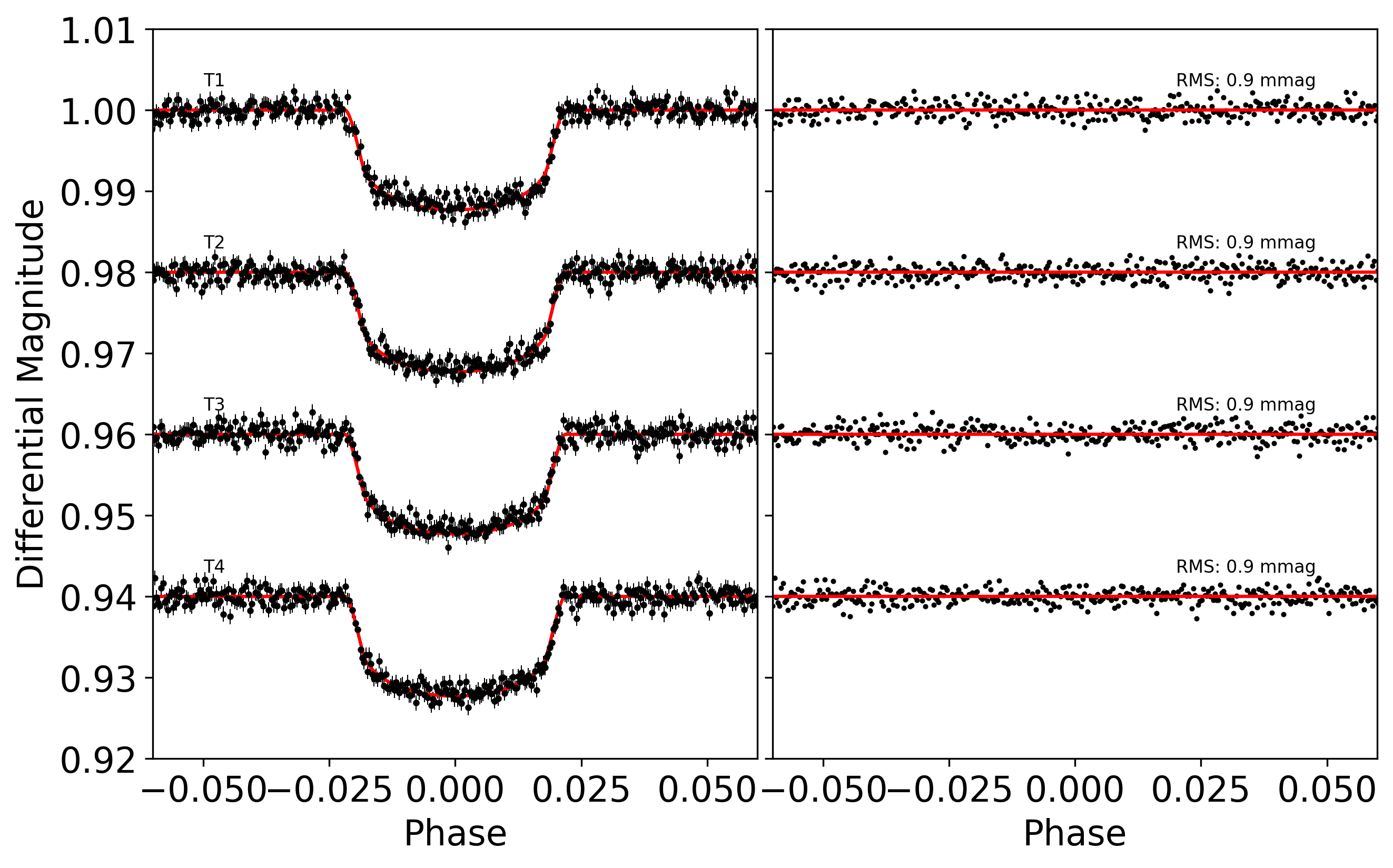}
    \caption{TESS observations of WASP-127\,b presented in this work. Left: detrended data and best-fit model. Right: residuals from fitting.}
    \label{fig:wasp127_tess}
\end{figure}

\begin{figure}
    \centering
    \includegraphics[width=\columnwidth]{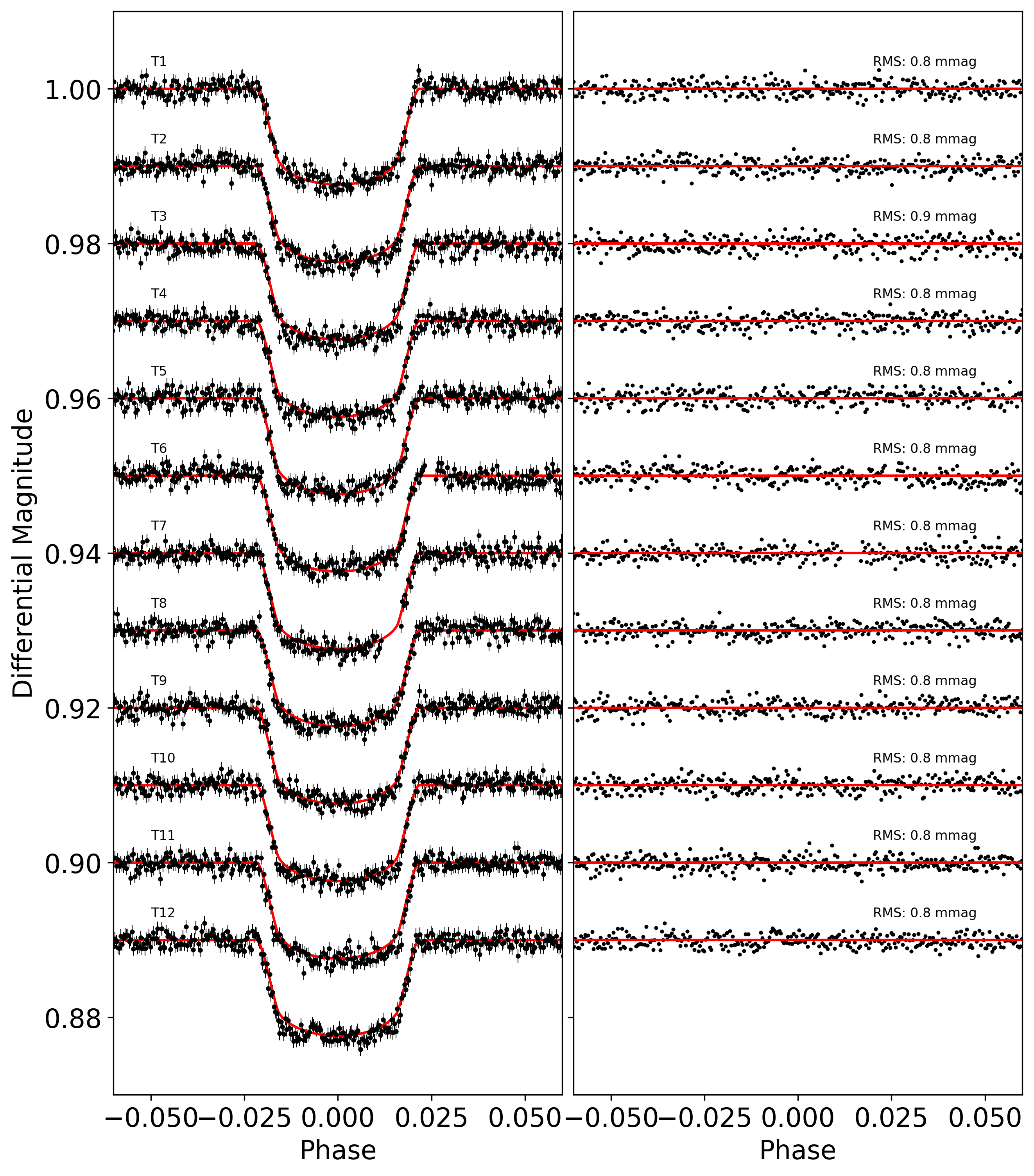}
    \caption{TESS observations of WASP-79\,b presented in this work. Left: detrended data and best-fit model. Right: residuals from fitting.}
    \label{fig:wasp79_tess}
\end{figure}

\begin{figure}
    \centering
    \includegraphics[width=\columnwidth]{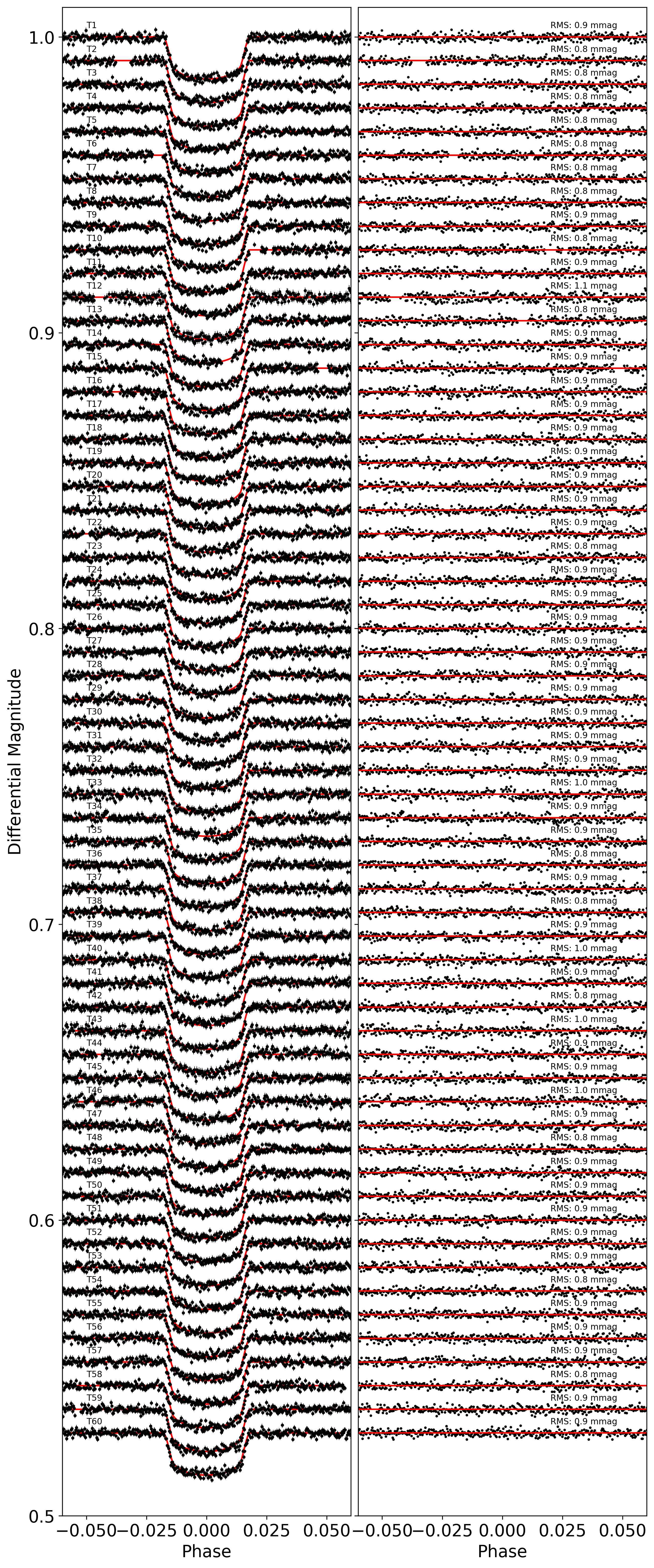}
    \caption{TESS observations of WASP-62\,b presented in this work. Left: detrended data and best-fit model. Right: residuals from fitting.}
    \label{fig:wasp62_tess}
\end{figure}

The white light curve fitting for the TESS data are in Figures \ref{fig:wasp127_tess} for WASP-127\,b, \ref{fig:wasp79_tess} for WASP-79\,b and \ref{fig:wasp62_tess} for WASP-62\,b.\\
All transit mid times used for the fitting are listed in Table \ref{tab:mid_times} and Table \ref{tab:mid_times2}.

{}

{}

{}

\begin{table}
    \centering
     \caption{Transit mid times used to refine the ephemeris of planets from this study. All mid times reported in this work are from TESS unless otherwise stated.}
     \resizebox{\columnwidth}{!}{
    \begin{tabular}{cccc}\hline \hline
     Planet & Epoch &  Mid Time [$BJD_{TDB}$] & Reference\\ \hline \hline
    WASP-127 b & -103 & 2457808.60283 $\pm$ 0.00031 & \cite{Palle} \\
    WASP-127 b & -5 & 2458218.053097 $\pm$ 0.000101 & This Work* \\
    WASP-127 b & 74 & 2458548.11973 $\pm$ 0.000469 & This Work \\
    WASP-127 b & 75 & 2458552.297636 $\pm$ 0.000428 & This Work \\
    WASP-127 b & 77 & 2458560.65431 $\pm$ 0.000435 & This Work \\
    WASP-127 b & 78 & 2458564.83219 $\pm$ 0.000431 & This Work \\ \hline
    WASP-79 b & -531 & 2456215.4556 $\pm$ 0.0005 & \cite{brown_rm} \\
    WASP-79 b & -94 & 2457815.92207 $\pm$ 0.000117 & This Work* \\
    WASP-79 b & 69 & 2458412.892172 $\pm$ 0.000299 & This Work \\
    WASP-79 b & 70 & 2458416.554571 $\pm$ 0.000311 & This Work \\
    WASP-79 b & 73 & 2458427.541703 $\pm$ 0.000332 & This Work \\
    WASP-79 b & 74 & 2458431.204093 $\pm$ 0.000347 & This Work \\
    WASP-79 b & 75 & 2458434.866496 $\pm$ 0.000321 & This Work \\
    WASP-79 b & 76 & 2458438.528882 $\pm$ 0.00029 & This Work \\
    WASP-79 b & 77 & 2458442.191242 $\pm$ 0.000292 & This Work \\
    WASP-79 b & 78 & 2458445.853639 $\pm$ 0.000326 & This Work \\
    WASP-79 b & 79 & 2458449.515994 $\pm$ 0.000314 & This Work \\
    WASP-79 b & 80 & 2458453.178366 $\pm$ 0.000317 & This Work \\
    WASP-79 b & 81 & 2458456.840324 $\pm$ 0.000363 & This Work \\
    WASP-79 b & 82 & 2458460.503155 $\pm$ 0.000335 & This Work \\\hline
    WASP-62 b & -614 & 2455767.1533 $\pm$ 0.0005 & \cite{brown_rm} \\
    WASP-62 b & -140 & 2457858.41397 $\pm$ 0.000266 & This Work* \\
    WASP-62 b & -34 & 2458326.078168 $\pm$ 0.000293 & This Work \\
    WASP-62 b & -33 & 2458330.490547 $\pm$ 0.000294 & This Work \\
    WASP-62 b & -32 & 2458334.902385 $\pm$ 0.000315 & This Work \\
    WASP-62 b & -30 & 2458343.725961 $\pm$ 0.00025 & This Work \\
    WASP-62 b & -28 & 2458352.550282 $\pm$ 0.000278 & This Work \\
     WASP-62 b & -27 & 2458356.961811 $\pm$ 0.000257 & This Work \\
    WASP-62 b & -26 & 2458361.374157 $\pm$ 0.000268 & This Work \\
    WASP-62 b & -25 & 2458365.786076 $\pm$ 0.000295 & This Work \\
    WASP-62 b & -24 & 2458370.197678 $\pm$ 0.000244 & This Work \\
    WASP-62 b & -23 & 2458374.610308 $\pm$ 0.000284 & This Work \\
    WASP-62 b & -22 & 2458379.021593 $\pm$ 0.00025 & This Work \\
    WASP-62 b & -20 & 2458387.845506 $\pm$ 0.000238 & This Work \\
    WASP-62 b & -19 & 2458392.257448 $\pm$ 0.000258 & This Work \\
    WASP-62 b & -17 & 2458401.081679 $\pm$ 0.000286 & This Work \\
    WASP-62 b & -16 & 2458405.493547 $\pm$ 0.000294 & This Work \\
    WASP-62 b & -14 & 2458414.317213 $\pm$ 0.000241 & This Work \\
    WASP-62 b & -12 & 2458423.141634 $\pm$ 0.000372 & This Work \\
    WASP-62 b & -11 & 2458427.55355 $\pm$ 0.0003 & This Work \\
    WASP-62 b & -10 & 2458431.96504 $\pm$ 0.000233 & This Work \\
    WASP-62 b & -9 & 2458436.376957 $\pm$ 0.000258 & This Work \\
    WASP-62 b & -1 & 2458471.672875 $\pm$ 0.000281 & This Work \\
    WASP-62 b & -0 & 2458476.084548 $\pm$ 0.000273 & This Work \\
    WASP-62 b & 1 & 2458480.496509 $\pm$ 0.000269 & This Work \\
    WASP-62 b & 2 & 2458484.90846 $\pm$ 0.000263 & This Work \\
    WASP-62 b & 3 & 2458489.320399 $\pm$ 0.000275 & This Work \\
    WASP-62 b & 4 & 2458493.732346 $\pm$ 0.000274 & This Work \\
    WASP-62 b & 5 & 2458498.144285 $\pm$ 0.000271 & This Work \\
    WASP-62 b & 6 & 2458502.556272 $\pm$ 0.000263 & This Work \\
    WASP-62 b & 7 & 2458506.968201 $\pm$ 0.000308 & This Work \\
    WASP-62 b & 8 & 2458511.380161 $\pm$ 0.000259 & This Work \\\hline \hline
    \multicolumn{4}{c}{*Data from Hubble}\\\hline \hline
    \end{tabular}
    }
    \label{tab:mid_times}
\end{table}

\begin{table}
    \centering
    \caption{Continuation of Table \ref{tab:mid_times}.}
    \resizebox{\columnwidth}{!}{
    \begin{tabular}{cccc}\hline \hline
     Planet & Epoch &  Mid Time [$BJD_{TDB}$] & Reference\\ \hline \hline
    WASP-62 b & 9 & 2458515.793014 $\pm$ 0.000308 & This Work \\
    WASP-62 b & 10 & 2458520.203407 $\pm$ 0.00031 & This Work \\
    WASP-62 b & 11 & 2458524.616007 $\pm$ 0.000263 & This Work \\
    WASP-62 b & 14 & 2458537.851688 $\pm$ 0.000269 & This Work \\
    WASP-62 b & 16 & 2458546.675279 $\pm$ 0.000324 & This Work \\
    WASP-62 b & 17 & 2458551.087738 $\pm$ 0.000337 & This Work \\
    WASP-62 b & 19 & 2458559.91167 $\pm$ 0.000278 & This Work \\
    WASP-62 b & 20 & 2458564.323603 $\pm$ 0.000285 & This Work \\
    WASP-62 b & 22 & 2458573.147116 $\pm$ 0.000311 & This Work \\
    WASP-62 b & 23 & 2458577.559142 $\pm$ 0.000317 & This Work \\
    WASP-62 b & 25 & 2458586.382966 $\pm$ 0.000333 & This Work \\
    WASP-62 b & 26 & 2458590.795328 $\pm$ 0.000345 & This Work \\
    WASP-62 b & 27 & 2458595.206831 $\pm$ 0.000331 & This Work \\
    WASP-62 b & 29 & 2458604.031175 $\pm$ 0.00029 & This Work \\
    WASP-62 b & 30 & 2458608.443135 $\pm$ 0.000301 & This Work \\
    WASP-62 b & 31 & 2458612.855056 $\pm$ 0.000311 & This Work \\
    WASP-62 b & 32 & 2458617.266322 $\pm$ 0.000359 & This Work \\
    WASP-62 b & 33 & 2458621.67842 $\pm$ 0.000344 & This Work \\
    WASP-62 b & 34 & 2458626.090956 $\pm$ 0.000358 & This Work \\
    WASP-62 b & 35 & 2458630.502921 $\pm$ 0.000265 & This Work \\
    WASP-62 b & 36 & 2458634.914867 $\pm$ 0.000257 & This Work \\
    WASP-62 b & 38 & 2458643.738117 $\pm$ 0.000318 & This Work \\
    WASP-62 b & 39 & 2458648.150206 $\pm$ 0.000305 & This Work \\
    WASP-62 b & 40 & 2458652.561559 $\pm$ 0.000312 & This Work \\
    WASP-62 b & 41 & 2458656.974299 $\pm$ 0.000292 & This Work \\
    WASP-62 b & 42 & 2458661.385862 $\pm$ 0.000292 & This Work \\
    WASP-62 b & 43 & 2458665.798175 $\pm$ 0.000285 & This Work \\
    WASP-62 b & 44 & 2458670.210186 $\pm$ 0.000304 & This Work \\
    WASP-62 b & 45 & 2458674.621865 $\pm$ 0.000315 & This Work \\
    WASP-62 b & 46 & 2458679.034395 $\pm$ 0.000284 & This Work \\\hline \hline
    \end{tabular}
    }
    \label{tab:mid_times2}
\end{table}

\bibliographystyle{aasjournal}
\bibliography{main}

\end{document}